\documentclass{aa}

\usepackage[]{graphicx}
\usepackage[]{exscale}

\newcommand{\mathsize}[1]{$ \displaystyle #1 $}

\def\aa#1#2#3{#1,      {A\&A, }{\bf#2}, #3}
\def\aas#1#2#3{#1,     {A\&AS, }{\bf#2}, #3}
\def\apj#1#2#3{#1,     {ApJ, }{\bf#2}, #3}

\def\apjsupp#1#2#3{#1, {ApJS, }{\bf#2}, #3}
\def\jcp#1#2#3{#1,     {J. Chem. Phys., }{\bf#2}, #3}
\def\rmp#1#2#3{#1,     {Rev. Mod. Physics., }{\bf#2}, #3}
\def\mn#1#2#3{#1,      {MNRAS, }{\bf#2}, #3}
\def\PSP#1#2#3{#1,     {Protostars and Planets IV, }{#2}, #3}
\def\pasj#1#2#3{#1,    {Publ. Astron. Soc. Japan, }{\bf#2}, #3}

\begin{document}

% title:
% ======
\title{On the Ionisation Fraction in
Protoplanetary Disks I: \\ Comparing Different Reaction Networks}

\author{Martin~Ilgner and Richard P. Nelson}

%\offprints{}

\institute{Astronomy Unit, Queen Mary, University of London, 
Mile End Road, London E1 4NS, U.K.}

\date{Received 22 June 2005 / Accepted 13 September 2005}

\titlerunning{Ionisation fraction in disks}

\authorrunning{M.~Ilgner, R.P.~Nelson}

% abstract:
% =========
\abstract{We calculate the ionisation fraction in protostellar disk models
using a number of different chemical reaction networks, including
gas--phase and gas--grain reaction schemes.
The disk models we consider are conventional
$\alpha$--disks, which include viscous heating and radiative cooling. 
The primary source of ionisation is assumed to be X--ray irradiation
from the central star. For most calculations we adopt a specific disk 
model (with accretion rate ${\dot M}=10^{-7}$ M$_{\odot}$yr$^{-1}$ 
and $\alpha=10_{}^{-2}$), and examine the predictions made by the 
chemical networks concerning the ionisation fraction, magnetic Reynolds 
number, and spatial extent of magnetically active regions. This is to 
aid comparison between the different chemical models. \\
We consider a number of gas--phase chemical networks. The simplest is
the five species model proposed by Oppenheimer \& Dalgarno (1974). We 
construct more complex models by extracting species and reactions from 
the UMIST data base. In general we find that the simple models predict 
higher fractional ionisation levels and more extensive active zones than 
the more complex models. When heavy metal atoms are included the simple 
models predict that the disk is magnetically active throughout. The 
complex models predict that extensive regions of the disk remain 
magnetically uncoupled (``dead'') even when the fractional abundance of 
magnesium $x_{\rm Mg}=10^{-8}$. This is because of the large number of 
molecular ions that are formed, which continue to dominate the 
recombination with free electrons in the presence of magnesium.\\
The addition of submicron sized grains with a concentration
of $x_{\rm gr}=10^{-12}$ causes the size of the ``dead zone'' to increase 
dramatically for all kinetic models considered, as the grains are highly 
efficient at sweeping up the free electrons. We find that the simple and 
complex gas--grain reaction schemes agree on the size and structure of the 
resulting ``dead zone'', as the grains play a dominant role in determining 
the ionisation fraction. We examine the effects of depleting the 
concentration of small grains as a crude means of modeling the growth of 
grains during planet formation. We find that a depletion factor of $10^{-4}$ 
causes the gas--grain chemistry to converge to the gas--phase chemistry when 
heavy metals are absent. When magnesium is included a depletion factor of 
$10^{-8}$ is required to reproduce the gas--phase ionisation fraction. This 
suggests that efficient grain growth and settling will be required in 
protoplanetary disks, before a substantial fraction of the disk mass in the 
planet forming zone between 1 -- 10 AU becomes magnetically active and 
turbulent. Only after this has occurred can gas--phase chemical models be used 
to predict reliably the ionisation degree in protoplanetary disks. 
% => keyword
}

\maketitle
%
%%%%%%%%%%%%%%%%%%%%%%%%%%%%%%%%%%%%%%%%%%%%%%%%%%%%%%%%%%%%%%%%%%%%
%
% section: Introduction
\section{Introduction}
Observations of young stars in star forming regions
point to the ubiquity of protostellar disks. These disks
show evidence for active accretion, with the canonical
mass flow rate being $10^{-8}$ M$_{\odot}$yr$^{-1}$
(e.g. Sicilia--Aguilar et al. 2004), requiring a means of
transporting angular momentum in the disks. 
At the present time only one mechanism has been shown to work:
MHD turbulence generated by the magnetorotational instability
(Balbus \& Hawley 1991; Hawley \& Balbus 1991; Hawley, Gammie \& 
Balbus 1996).\\
\indent
There are questions, however, about the applicability of the MRI 
to cool and dense protostellar disks, as the ionisation fraction
is expected to be low (e.g. Blaes \& Balbus 1994; Gammie 1996). 
Magnetohydrodynamic simulations of disks including ohmic resistivity 
(Fleming, Stone \& Hawley 2000) show that for magnetic Reynolds 
numbers ${Re}_{\rm m}^{}$ smaller than a critical value 
${Re}_{\rm m}^{\rm crit}$, turbulence cannot be sustained and the disks 
return to a near--laminar state. The value of ${Re}_{\rm m}^{\rm crit}$ 
depends on the magnetic field configuration. For net--flux vertical 
fields turbulence is sustained when ${Re}_{\rm m}^{} \ge 100$. For 
zero--net flux fields, turbulence dies when ${Re}_{\rm m}^{} < 10^4$. 
Although there remains some uncertainty about the precise value of 
${Re}_{\rm m}^{\rm crit}$ due to the possible role of additional 
non--ideal MHD effects, recent simulations by Sano \& Stone (2002) 
indicate that the inclusion of Hall E.M.F.s has little impact on the 
value of ${Re}_{\rm m}^{\rm crit}$. A value of ${Re}_{\rm m}^{\rm crit}=100$ 
typically corresponds to a gas--phase electron fraction of 
$x[\rm e_{}^{-}] \simeq 10^{-12}$ for most disk models.\\
\indent
The question of whether protostellar disks can sustain MHD turbulence 
has important implications beyond the issue of how mass accretes onto 
the central protostar. In particular turbulence may play an important 
role in planet formation. The growth of planetesimals is controlled by 
the velocity dispersion (e.g. Wetherill \& Stewart 1993), as is the 
growth of protoplanetary cores (Kokubo \& Ida 2000; Thommes et al. 2003). 
Simulations of planets in disks indicate that turbulence may have a 
significant influence on planet migration and gap formation, as well as 
the velocity dispersion of planetesimals (Nelson \& Papaloizou 2003, 2004; 
Winters, Balbus \& Hawley 2003; Nelson 2005). Obtaining detailed knowledge 
of the structure of turbulent protoplanetary disks is a crucial part of 
developing models of planet formation.\\
\indent
A number of studies of the ionisation fraction in protostellar disks have 
appeared in the literature. The early work by Gammie (1996) put forward the 
idea that disks may have magnetically ``active zones" sustained by thermal 
or cosmic ray ionisation, adjoining regions that were magnetically inactive 
-- ``dead zones". More recent studies have examined this issue, but in each 
case different chemical reaction networks and/or disk models have been 
employed, making comparison between them difficult. For example, Gammie (1996) 
used a layered $\alpha$--disk model, and considered cosmic--rays as the 
primary ionising source. Sano et al. (2000) used the Hayashi (1981) minimum 
mass solar nebula model in conjunction with a more complex chemical model that 
included dust grains. Glassgold et al. (1997) and Igea et al. (1999) used the 
minimum mass model and considered X--rays as the most important ionisation 
source. Fromang et al. (2002) considered the ionisation fraction in 
conventional $\alpha$--disk models, and focussed on the role of heavy metals in 
a simple reaction network. Matsumura \& Pudritz (2003) adopted the externally 
heated, passive disk model proposed by Chiang \& Goldreich (1997) in conjunction 
with the Sano et al. (2000) chemical reaction network. Recent work by Semenov et 
al. (2004) studied chemistry in the D'Alessio et al. (1999) disk model, and 
employed a complex gas--grain chemical network including most reaction from the 
UMIST data base.\\
\indent
In this work we study the chemistry and ionisation fraction in conventional 
$\alpha$--disk models, similar to the approach taken by Fromang et al. (2002). 
We include X--ray ionisation from the central star as the primary (external) 
ionisation source, and neglect cosmic--ray ionisation. One of our primary goals 
is to compare the ionisation fraction predicted by the different reaction 
networks, within the framework of the same underlying disk model.\\
\indent
We study both gas--phase chemical models, and gas--grain models. We begin by 
examining the ionisation fraction generated by the simple gas--phase, five 
species model proposed by Oppenheimer \& Dalgarno (1974) for studying the 
electron fraction in dark clouds. We study also a number of more complex 
gas--phase chemical networks. These include the Sano et al. (2000) kinetic model 
(applied to the gas phase), and networks constructed from various species sets 
that are drawn from the UMIST data base (Le Teuff et al. 2000). In general we 
find disagreement between these models, with the more complex networks in 
particular producing lower fractional ionisation than the simpler schemes. This 
is because a large number of molecular ions form that are able to recombine 
quickly with free electrons. Even a relatively large abundance of heavy metals 
(magnesium) is unable to render the disks fully active in these cases.\\
\indent
We construct a series of gas--grain chemical models that are generalisations of 
the gas--phase networks. In line with expectations, we find that the addition of 
submicron sized ($r_{\rm gr}^{}=10^{-5}$ cm) grains causes the magnetically active 
zone to shrink dramatically, with each of the models being in basic agreement about 
its size and structure. As a simple approach to modelling the effects of growth 
during planet formation, we examine the effects of depleting small grains on the 
chemistry. We find that depletion factors of $10^{-4}$ are required to reproduce the 
gas--phase chemistry for models in which heavy metals are absent. Depletion factors 
of $\simeq 10^{-8}$, however, are required to reproduce the ionisation fraction 
obtained when metals are included. The effectiveness of very small numbers of grains 
in modifying the equilibrium ionisation fraction depends crucially on there being a 
constant supply of neutral grains, these being generated through recombination 
reactions between negatively charged grains and positive ions. These models suggest 
that efficient growth of small grains will be required before they can safely be 
neglected from chemical networks that predict the ionisation fraction in 
protoplanetary disks.\\
\indent
This paper is organised as follows. The $\alpha$--disk models we construct are 
described in section~\ref{disk_models}. The chemical models are described in 
section~\ref{chemical_models}, and our scheme for calculating the X--ray ionisation 
rate is described in section~\ref{free_electrons}. The results of the calculations 
are presented in section~\ref{results}, and extensions to the basic models are 
presented in section~\ref{discussion}. Finally we summarise the paper in 
section~\ref{summary}.

%%%%%%%%%%%%%%%%%%%%%%%%%%%%%%%%%%%%%%%%%%%%%%%%%%%%%%%%%%%%%%%%%%%%
\section{Disk models}
\label{disk_models}
\begin{figure}[t]
\includegraphics[width=.45\textwidth]{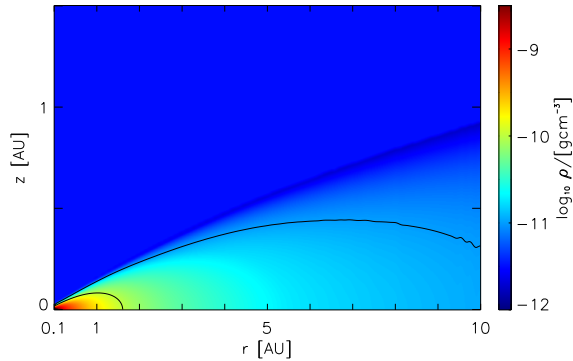}
\caption{This contour plot shows the density distribution for
our $\alpha=10^{-2}$, ${\dot M}=10^{-7}$ M$_{\odot}$yr$^{-1}$ 
disk model. The contour lines refer to values of $10^{-11}$ and 
$10^{-10} \ \mathrm{g cm^{-3}}$, respectively.}
\label{figure_density}
\end{figure}

\begin{figure}[t]
\includegraphics[width=.45\textwidth]{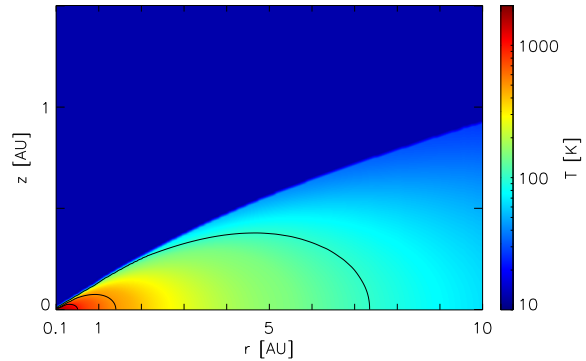}
\caption{This contour plot shows the temperature distribution for
our $\alpha=10^{-2}$, ${\dot M}=10^{-7}$ M$_{\odot}$yr$^{-1}$ 
disk model. The contour lines refer to values of 100, 500, and 
1000 K, respectively.}
\label{figure_temp}
\end{figure}

Numerous studies of the ionisation structure in protoplanetary
disks have appeared in the literature, with a variety of disk 
models being used. The resulting ionisation structure differs 
from model to model, as they differ in their density and temperature 
profiles. The early study by Gammie (1996) used a layered 
$\alpha$--disk model. Sano et al. (2000) used the Hayashi (1981)
minimum mass solar nebula model, as did Glassgold et al. 
(1997) and Igea et al. (1999). Matsumura \& Pudritz (2003) adopted 
the externally heated, passive disk model proposed by Chiang \& 
Goldreich (1997), and recent work by Semenov et al. (2004) studied 
chemistry in the D'Alessio et al. (1999) disk model which includes 
U.V. heating from the central star and internal viscous 
heating. In this work we study the chemistry and ionisation fraction 
in conventional $\alpha$--disk models, similar to the approach taken 
by Fromang et al. (2002).\\
\indent
A detailed description of the disk model and modelling procedure is 
given in Ilgner et al. (2004). Here we just sketch out the important 
details. We compute a disk model with inner and outer radii [0.1, 10] 
AU orbiting with Keplerian velocity about a central star with mass 1 
M$_{\odot}$. The vertical structure is obtained at a given radius by 
solving for hydrostatic and thermal equilibrium. Radiative and convective 
transport is included with opacities taken from Semenov et al. (2003), 
and disk heating occurs through viscous dissipation alone. With these 
ingredients and appropriate boundary conditions the disk model is 
uniquely specified for given values of $\alpha$ and the mass accretion 
rate $\dot M$. Most of our models are computed with $\alpha=10^{-2}$ and 
${\dot M}=10^{-7}$ M$_{\odot}$yr$^{-1}$, giving a mass of 0.0087 
M$_{\odot}$ within the computational domain. Contour plots showing the 
density and temperature distributions are shown in 
figures~\ref{figure_density} and \ref{figure_temp}. We consider a model 
with $\alpha=5 \times 10^{-3}$ and 
${\dot M}=10^{-8}$ M$_{\odot}$yr$^{-1}$ toward the end of this paper. 
Similar models and procedures for calculating $\alpha$--models were used 
by Papaloizou \& Terquem (1999) and Fromang et al. (2002).
% ================================================================= 
% figure 1: density profile 
% =================================================================
% =================================================================

%%%%%%%%%%%%%%%%%%%%%%%%%%%%%%%%%%%%%%%%%%%%%%%%%%%%%%%%%%%%%%%%%%%%
\section{Chemical models}
\label{chemical_models}
We use standard techniques to solve the kinetic equations 
that describe the chemical models we consider in this paper. 
We aim to provide sufficient detail to enable an interested 
reader to reproduce our results, so full descriptions of our 
procedures are presented in appendices. In this section we begin 
by describing important aspects of the kinetic models we have 
implemented, prior to discussing rate coefficients and initial 
abundances.

%%%%%%%%%%%%%%%%%%%%%%%%%%%%%%%%%%%%%%%%%%%%%%%%%%%%%%%%%%%%%%%%%%%%
\subsection{Kinetic models}
A \emph{kinetic model} is defined by the \emph{components} (or 
species), the \emph{chemical reactions} that cause time dependent 
changes in the abundances of species, and by the underlying 
\emph{kinetic equation} for each component. The kinetic equations 
are given by a set of stiff ordinary differential equations which 
are solved numerically. We have confirmed both algebraically and 
numerically that conservation of elements and charge is satisfied 
by our kinetic models. We have implemented the following reaction 
schemes, and examined the resulting ionisation fractions in 
$\alpha$--disk models.
%%%%%%%%%%%%%%%%%%%%%%%%%%%%%%%%%%%%%%%%%%%%%%%%%%%%%%%%%%%%%%%%%%%%
% Model by Oppenheimer and Dalgarno:  
\subsubsection{Oppenheimer \& Dalgarno model}
Oppenheimer \& Dalgarno (1974) introduced a simplified reaction 
scheme to approximate the equilibrium electron abundance in dense 
% ================================================================= 
% figure 2: temperature profile 
% =================================================================
% =================================================================
interstellar clouds. Using an analytic approximation, this scheme 
has been used to calculate the electron fraction in protoplanetary 
disks by a number of authors (Gammie 1996; Glassgold et al. 1997; 
Fromang et al. 2002).\\
\indent
For a given total particle density $n$, Oppenheimer \& Dalgarno 
introduced a simple kinetic model involving two elements (``m" and ``M"), 
five components (``m", ``m$_{}^+$", ``M", ``M$_{}^+$", and 
``e$_{}^-$"), and four chemical reactions. Here, m and m$_{}^+$ 
represents a molecule and its ionized counterpart, while M and M$_{}^+$ 
refer to a heavy neutral metal atom and its ionized counterpart.
e$_{}^-$ denotes the free electron in the gas phase.\\
\indent
In this model, ions are produced by ionisation and a charge--transfer 
reaction, while dissociative recombination and radiative recombination 
are assumed to destroy ions. The reaction set is represented 
schematically in table~\ref{list_oppenheimer}, where $\zeta$, 
$\tilde\alpha$, $\tilde\beta$, and $\tilde\gamma$ are the ionizing flux, 
the dissociative recombination rate coefficient, the charge--transfer rate 
coefficient, and the radiative recombination rate coefficient.\\
\indent
One can derive the equilibrium electron abundance analytically using a 
simple approximation (which is $n \approx n[\mathrm{m}]$ where 
$n[\mathrm{m}]$ denotes the number density of the neutral molecule m). 
The equilibrium electron concentration $x_{\infty}[\mathrm{e}^{-}]$ is 
given by the roots of the algebraic equation\footnote{For simplicity we 
dropped the symbol $\infty$ from $x[\mathrm{e}^{-}]$ and $x[\mathrm{M}]$.}
% =================================================================
% algebraic equation, Oppenheimer & Dalgarno
% =================================================================
\begin{equation}
x[\mathrm{e}_{}^{-}]^3 + 
\frac{\tilde{\beta} x[\mathrm{M}]}{\tilde{\alpha}} \ x[\mathrm{e}_{}^{-}]^2 -
\frac{\zeta}{\tilde{\alpha} n} \ x[\mathrm{e}_{}^{-}] - 
\frac{\tilde{\beta}\zeta x[\mathrm{M}]}{\tilde{\alpha}\tilde{\gamma} n}  
= 0, \
\label{eq_oppenheimer} 
\end{equation}
% =================================================================
\indent
where $x[\mathrm{M}]$ is the equilibrium particle 
% =================================================================
% table: Oppenheimer & Dalgarno
% =================================================================
\begin{table}[t]
\caption{List of reactions (and the corresponding rate 
coefficients) considered by Oppenheimer \& Dalgarno (1974).}
\begin{center}
\begin{tabular}{l@{\extracolsep{0.1cm}}l@{\extracolsep{0.1cm}}
c@{\extracolsep{0.2cm}}l@{\extracolsep{0.2cm}}
c@{\extracolsep{0.2cm}}l@{\extracolsep{0.2cm}}
c@{\extracolsep{0.2cm}}l@{\extracolsep{0.2cm}}
l@{\extracolsep{0.2cm}}}\hline \hline
1. & m$_{}^{}$          &   &            
 & $\longrightarrow$   & 
    m$_{}^{+}$ & + & e$_{}^{-}$   & $\zeta$  \\  
2. & m$_{}^{+}$ & + & e$_{}^{-}$ 
 & $\longrightarrow$  & 
    m$_{}^{}$      &   &          & $\tilde{\alpha} = 3 \times 10^{-6} / \sqrt{T} \ \rm cm_{}^3s_{}^{-1}$  \\
3. & M$_{}^{+}$ & + & e$_{}^{-}$ 
 & $\longrightarrow$  & 
    M$_{}^{}$      & + &  $h\nu$  & $\tilde{\gamma} = 3 \times 10^{-11} / \sqrt{T} \ \rm cm_{}^3s_{}^{-1}$ \\ 
4. & m$_{}^{+}$ & + & M$_{}^{}$  
 & $\longrightarrow$   & 
    m$_{}^{}$      & + & M$_{}^{+}$  & $\tilde{\beta} = 3 \times 10^{-9} \ \rm cm_{}^3s_{}^{-1}$  \\ \hline
\end{tabular}
\end{center}
\label{list_oppenheimer}
\end{table}
% =================================================================
concentration of metals. Neglecting the contributions due to the heavy 
metal atom equation (\ref{eq_oppenheimer}) can be reduced to its 
quadratic form whose solution is
% =================================================================
% special solution, Oppenheimer & Dalgarno 
% =================================================================
\begin{equation}
x_{\infty}[\mathrm{e}_{}^{-}] = 
\sqrt{\frac{\zeta}{\tilde{\alpha} n}} \hspace*{.5cm} \mathrm{if} 
\ x_{\mathrm{M}}^{} = 0 \ .
\label{eq_oppenheimer2}
\end{equation}
% =================================================================
In the opposite limit, i.e. assuming 
$x_{\infty}[\mathrm{M}] \gg x_{\infty}[\mathrm{e}_{}^{-}]$, similar 
expressions can be derived depending on further assumption being made. 
See for example equation (27) in Oppenheimer \& Dalgarno (1974) and 
equation (14) in Fromang et al. (2002).

%%%%%%%%%%%%%%%%%%%%%%%%%%%%%%%%%%%%%%%%%%%%%%%%%%%%%%%%%%%%%%%%%%%%
\subsubsection{Umebayashi \& Nakano model}
Sano et al. (2000) calculated the ionisation fraction in a minimum 
mass solar nebula disk model adopting a reaction scheme which was 
investigated originally by Umebayashi \& Nakano (1990) and Nishi et 
al. (1991) under dense interstellar cloud conditions.\\
\indent
This reaction scheme extends the more schematic gas--phase scheme of 
Oppenheimer \& Dalgarno (1974) by introducing additional neutral and 
ionized gas--phase species, and gas--grain interactions. Sano et 
al. (2000) considered 7 ions (H$_{}^+$, H$^+_2$, H$^+_3$, 
He$_{}^+$, C$_{}^+$, m$_{}^+$, M$_{}^+$) and grain particles with 
seven different grain charges (gr, gr$_{}^{\pm}$, gr$_{}^{2\pm}$,
gr$_{}^{3\pm}$). The ions m$_{}^+$ and M$_{}^+$ represent a molecular 
and metal ion, respectively.\\
\indent
The reactions considered by Sano et al. (2000) are listed in 
table~\ref{table_sano} which is taken from their paper. Note 
that the ions O$_{}^+$, O$_2^+$, OH$_{}^+$, O$_2^{}$H$_{}^+$, 
CO$_{}^+$, CH$_2^+$, and HCO$_{}^+$ are represented by the molecular 
ion m$_{}^+$ (Sano, private communication). Our only omission 
was triple charged grains, which Sano et al. (2000) included.\\
\indent
The results obtained by Sano et al. (2000) can be used for testing 
our implementation of this kinetic model. Applying the same physical 
% =================================================================
% Table: Sano
% =================================================================
\begin{table}[hd]
\caption{List of reactions considered by Sano et al. (2000) excluding 
triple charged grains. Notation: X$^+_{}$ denotes the ions of 
the kinetic model: [H$^{+}_{}$, H$^{+}_{2}$, H$^{+}_{3}$, He$^{+}_{}$, 
C$^{+}_{}$, m$^{+}_{}$, M$^{+}_{}$] while Y, Y$_{\rm ads}$ 
refer to neutral gas--phase species and adsorbed species onto grain 
particles, respectively.}
\begin{center}
\begin{tabular}{%
%|rlclclclcll|}\hline
 r@{\extracolsep{0.1cm}}l@{\extracolsep{0.1cm}}% 1. a
 c@{\extracolsep{0.1cm}}l@{\extracolsep{0.1cm}}%      + b
 c@{\extracolsep{0.3cm}}l@{\extracolsep{0.1cm}}%           -> c
 c@{\extracolsep{0.2cm}}l@{\extracolsep{0.1cm}}%                + d
% c@{\extracolsep{0.1cm}}l|@{\extracolsep{0.1cm}}
% @{\extracolsep{0.1cm}}l|@{\extracolsep{0.1cm}}}\hline
 c@{\extracolsep{0.1cm}}l}\hline\hline
1. & H$_{2}^{ }$              &   &      
   & $\longrightarrow$  
   & H$_{2}^{+}$              & + &  e$_{}^{-}$  &   &
%   & 0.97$\zeta$ 
\\
2. & H$_{2}^{}$               &   &   
   & $\longrightarrow$  
   & H$_{}^{+}$               & + &   H          & + &  e$_{}^{-}$
%   & 0.03$\zeta$ 
\\
3. & He                       &   &    
   & $\longrightarrow$  
   & He$_{}^{+}$              & + &   e$_{}^{-}$ &   &
%   & 0.84$\zeta$ 
\\
4. & H$_{}^{+}$               & + &   Mg  
   & $\longrightarrow$  
   & Mg$_{}^{+}$              & + &   H          &   &
%   &  
\\
5. & H$_{}^{+}$               & + &   O$_{}^{}$   
   & $\longrightarrow$  
   & O$_{}^{+}$               & + &   H           &   &
%   &  
\\
6. & H$_{}^{+}$               & + &   O$_{2}^{}$  
   & $\longrightarrow$  
   & O$_{2}^{+}$              & + &   H           &   &
%   &  
\\
7. & H$_{2}^{+}$              & + &   H$_{2}^{}$  
   & $\longrightarrow$  
   & H$_{3}^{+}$              & + &   H           &   &
%   &  
\\   
8. & H$_{3}^{+}$              & + &   O   
   & $\longrightarrow$  
   & OH$_{}^{+}$              & + &   H$_{2}^{}$  &   &        
%   &  
\\
9. & H$_{3}^{+}$              & + &   Mg   
   & $\longrightarrow$  
   & Mg$_{}^{+}$              & + &   H$_{2}^{}$  & + & H           
%   &  
\\      
10.& H$_{3}^{+}$              & + &   CO  
   & $\longrightarrow$  
   & HCO$_{}^{+}$             & + &   H$_{2}^{}$  &   &         
%   &  
\\
11.& H$_{3}^{+}$              & + &   O$_{2}^{}$
   & $\longrightarrow$  
   & O$_{2}^{}$H$_{}^{+}$     & + &   H$_{2}^{}$  &   &           
%   &  
\\
12.& He$_{}^{+}$              & + &   H$_{2}^{}$  
   & $\longrightarrow$  
   & H$_{}^{+}$               & + &   H           & + & He        
%   &  
\\   
13.& He$_{}^{+}$              & + &   CO  
   & $\longrightarrow$  
   & C$_{}^{+}$               & + &   O           & + & He        
%   &  
\\
14.& He$_{}^{+}$              & + &   O$_{2}^{}$  
   & $\longrightarrow$  
   & O$_{}^{+}$               & + &   O           & + & He        
%   &  
\\
15.& C$_{}^{+}$               & + &   Mg  
   & $\longrightarrow$  
   & Mg$_{}^{+}$              & + &   C           &   &         
%   &  
\\
16.& C$_{}^{+}$               & + &   H$_{2}^{}$  
   & $\longrightarrow$  
   & CH$_{2}^{+}$             & + &   $h\nu$      &   &         
%   &  
\\
17.& C$_{}^{+}$               & + &   O$_{2}^{}$  
   & $\longrightarrow$  
   & CO$_{}^{+}$              & + &   O           &   &         
%   &  
\\ 
18.& C$_{}^{+}$               & + &   O$_{2}^{}$  
   & $\longrightarrow$  
   & O$_{}^{+}$               & + &   CO          &   &         
%   &  
\\
19.& HCO$_{}^{+}$             & + &   Mg$_{}^{}$  
   & $\longrightarrow$  
   & Mg$_{}^{+}$              & + &   HCO         &   &         
%   &  
\\
20.& H$_{}^{+}$               & + &   e$_{}^{-}$  
   & $\longrightarrow$  
   & H                        & + &   $h\nu$      &   &         
%   &  
\\
21.& H$_{3}^{+}$              & + &   e$_{}^{-}$  
   & $\longrightarrow$  
   & H$_{2}^{}$               & + &   H           &   &         
%   &  
\\
22.& H$_{3}^{+}$              & + &   e$_{}^{-}$  
   & $\longrightarrow$  
   & H$_{}^{}$                & + &   H           & + &  H        
%   &  
\\
23.& He$_{}^{+}$              & + &   e$_{}^{-}$  
   & $\longrightarrow$  
   & He$_{}^{}$               & + &   $h\nu$      &   &         
%   &  
\\
24.& C$_{}^{+}$               & + &   e$_{}^{-}$  
   & $\longrightarrow$  
   & C                        & + &   $h\nu$      &   &         
%   &  
\\
25.& Mg$_{}^{+}$              & + &   e$_{}^{-}$  
   & $\longrightarrow$  
   & Mg                       & + &   $h\nu$      &   &         
%   &  
\\
26.& HCO$_{}^{+}$             & + &   e$_{}^{-}$  
   & $\longrightarrow$  
   & CO                       & + &   H           &   &         
%   &  
\\
27.& X$_{}^{+}$               & + &   gr$_{}^{2-}$  
   & $\longrightarrow$  
   & gr$_{}^{-}$              & + &   Y           &   &         
%   &  
\\
28.& X$_{}^{+}$               & + &   gr$_{}^{-}$ 
   & $\longrightarrow$  
   & gr                       & + &   Y           &   &         
%   &  
\\     
29.& X$_{}^{+}$               & + &   gr$_{}^{+}$  
   & $\longrightarrow$  
   & gr$_{}^{2+}$             & + &   Y$_{\mathrm{ads}}^{}$ &   &         
%   &  
\\  
30.& X$_{}^{+}$               & + &   gr  
   & $\longrightarrow$  
   & gr$_{}^{+}$              & + &   Y$_{\mathrm{ads}}^{}$ &   &         
%   &  
\\ 
31.& e$_{}^{-}$               & + &   gr$_{}^{-}$  
   & $\longrightarrow$  
   & gr$_{}^{2-}$             &   &               &   &         
%   &  
\\ 
32.& e$_{}^{-}$               & + &   gr  
   & $\longrightarrow$  
   & gr$_{}^{-}$              &   &               &   &         
%   &  
\\
33.& e$_{}^{-}$               & + &   gr$_{}^{+}$  
   & $\longrightarrow$  
   & gr                       &   &               &   &         
%   &  
\\   
34.& e$_{}^{-}$               & + &   gr$_{}^{2+}$  
   & $\longrightarrow$  
   & gr$_{}^{+}$              &   &               &   &         
%   &  
\\
35.& gr$_{}^{+}$              & + &   gr${}^{-}$  
   & $\longrightarrow$  
   & gr                       & + &   gr          &   &         
%   &  
\\
36.& gr$_{}^{2+}$             & + &   gr$_{}^{2-}$  
   & $\longrightarrow$  
   & gr                       & + &   gr           &   &         
%   &  
\\
37.& gr$_{}^{+}$              & + &   gr$_{}^{2-}$  
   & $\longrightarrow$  
   & gr$_{}^{-}$              & + &   gr           &   &         
%   &  
\\
38.& gr$_{}^{-}$              & + &   gr$_{}^{2+}$  
   & $\longrightarrow$  
   & gr$_{}^{+}$              & + &   gr           &   &         
%   &  
\\  \hline  
\end{tabular}
\end{center}
\label{table_sano}
\end{table}%
% =================================================================
% =================================================================
% figure 3: Sano 
% =================================================================
\begin{figure}[t]
\includegraphics[width=.45\textwidth]{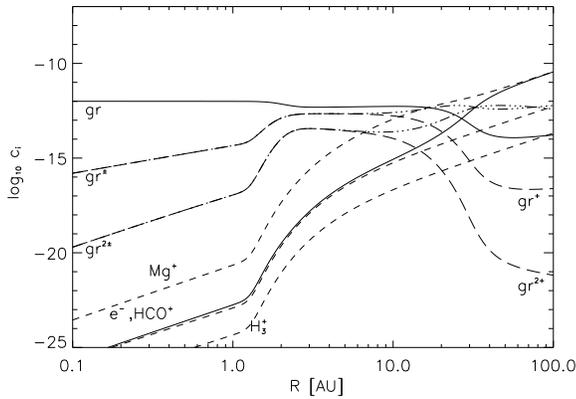}
\caption{Equilibrium abundances (per hydrogen nucleus) of some 
representative particles at the disk midplane, which should be
compared to the fiducial model of {\sc Sano} et al. (2000). The 
distribution of $\mathrm{e^-}$ and of $\mathrm{HCO^+}$ 
is given by the solid and the dashed line, respectively.}
\label{figure_sano_1}
\end{figure}
% =================================================================
conditions (which includes a surface density distribution $\Sigma(R)$ 
based on the minimum mass solar nebula model), the same ionisation 
flux, and rate coefficients (including the value of 
$S_{\mathrm{e}^{-}} = 0.6$ for all temperatures and grain charges, 
where $S_{\mathrm{e}^{-}}$ denotes the sticking probability of 
electrons onto neutral or charged grains), and the same equilibrium 
values of the abundances of neutral components\footnote{We assumed a 
value of $c_{\mathrm{gr}} = 10^{-12}$ for the concentration of grain 
particles per hydrogen nuclei. $r_{\mathrm{gr}} = 0.1 \ \mu m$ is 
taken as the radius of the grain particle.} as Sano et al. (2000) used, 
we are able to reproduce their results. The equilibrium distributions 
of the most important components are shown in figure~\ref{figure_sano_1} 
and they correspond extremely well to the fiducial model presented in 
figure~(1) of Sano et al. (2000). This indicates that our implementation 
of the Umebayashi \& Nakano reaction scheme is correct.

%%%%%%%%%%%%%%%%%%%%%%%%%%%%%%%%%%%%%%%%%%%%%%%%%%%%%%%%%%%%%%%%%%%%
\subsubsection{Extended kinetic models}
\label{extended_models}
The kinetic models introduced by Oppenheimer \& Dalgarno and by 
Umebayashi \& Nakano may serve as \emph{prototypes} for gas--phase 
chemistry and gas--grain chemistry, respectively. Based on these 
prototypes we have developed more complex kinetic models by extending 
the number of elements, species, and reactions.\\
\indent
We have formalised our approach to developing extended kinetic models 
as follows. We start with all gas--phase species listed in 
table~\ref{list_species_set}. 
This species set contains the elements
H, He, C, O, N, S, Si, Mg and Fe. We extract
reactions involving these species from the UMIST database
(Le Teuff et al. 2000). The same species set was used by 
Willacy et al. (1998), Willacy \& Langer (2000), and 
Markwick et al. (2002) in their studies of disk chemistry. 
This forms the basis of the gas--phase 
chemical network that we implement.\\
\indent
We also consider chemical evolution due to gas--grain collisions and 
grain--grain collisions. As yet there is no UMIST--like database that 
describes these types of reactions, so we have developed our own model. 
We use a nomenclature that differentiates between reactions that involve 
mantle species and those that do not. We refer to the former as 
``mantle chemistry" and the latter as ``grain chemistry". We now describe 
the mantle chemistry, followed by the grain chemistry.\\
\indent
Consider first the neutral components of the previously described 
gas--phase chemistry. We define a new class of species: {\em mantle species} 
-- which are the adsorbed counterparts of the gas--phase neutral species, 
with the exception of Helium which is assumed to remain in the gas phase. 
These reactions are represented schematically in 
table~\ref{list_mantle_chemistry}, where ``X" is a neutral species and the 
``gr" are grain particles whose charges are denoted by their indices. A 
collision between a neutral species X and a grain with charge $q$ simply 
% =================================================================
% Table: mantle chemistry
% =================================================================
\begin{table}[t]
\caption{List of generalised reactions determining the 
mantle chemistry. Neutral gaseous species are represented by 
the symbol X, its ionised counterpart by X$_{}^+$. The notation
$\mathrm{X(gr)}$ and $\mathrm{X(gr^{-})}$ refer to 
neutral gas--phase species adsorbed onto grain particles with 
no excess charge and with single negative excess charge, 
respectively.}
\begin{center}
\begin{tabular}{llclcl}\hline\hline
1. & X$_{}^{}$                & + &   gr$_{}^{2-}$  
   & $\longrightarrow$  
   & X(gr$_{}^{2-}$)          \\
2. & X$_{}^{}$                & + &   gr$_{}^{-}$ 
   & $\longrightarrow$  
   & X(gr$_{}^{-}$)           \\     
3. & X$_{}^{}$                & + &   gr$_{}^{}$  
   & $\longrightarrow$  
   & X(gr)                    \\  
4. & X$_{}^{}$                & + &   gr$_{}^{+}$  
   & $\longrightarrow$  
   & X(gr$_{}^{+}$)           \\ 
5. & X$_{}^{}$                & + &   gr$_{}^{2+}$  
   & $\longrightarrow$  
   & X(gr$_{}^{2+}$)          \\ 
6. & X(gr,gr$_{}^{\pm}$,gr$_{}^{2\pm}$)                       &   &     
   & $\longrightarrow$  
   & X                        \\  
7. & X$_{}^{+}$               & + &   gr$_{}^{+}$  
   & $\longrightarrow$  
   & X(gr$_{}^{2+}$)          \\ 
8. & X$_{}^{+}$               & + &   gr  
   & $\longrightarrow$  
   & X(gr$_{}^{+}$)           \\ \hline   
\end{tabular}
\end{center}
\label{list_mantle_chemistry}
\end{table}%
% =================================================================
results in the formation of a mantle species ${\rm X}({\rm gr}^q)$ which 
physically represents atoms/molecules of species X sitting on grains with 
charge $q$.\\
\indent
We now consider the ``grain chemistry". The reactions are represented 
schematically in table~\ref{list_grain_chemistry}. Reactions 1 and 2 show 
that a positive ion colliding with a negative grain results in the ion being 
neutralised, but remaining in the gas phase, and the grain being left with 
one less charge -- provided a neutral counterpart X of the ion X$^+$ exists 
in the UMIST database. Collisions between electrons and grains neutralise 
positive grains, and negatively charge neutral or singly--negative charged 
grains. The model incorporates single and double charging of grains. 
Grain--grain collisions result in conservation of grain number but involve
charge exchange. There are some species in the set described in 
table~\ref{list_species_set} whose collisions with grains cannot be described 
by these reactions. In particular these are collisions between grains and 
those ions X$^+$ (such as H$_3^+$ and NH$^+_4$) with no neutral counterpart. 
The products of these collisions are unknown. By neglecting these particular 
collisions between ions and grains one may overestimate the ionisation 
fraction. To remedy this we have assumed that collisions between ions X$^+$ 
and negatively charged grains form the same products as the dissociative 
reactions in the gas phase:
% =================================================================
% table:
% =================================================================
\begin{center}
\begin{tabular}{lclclclcl}
X$_{}^{+}$  & + &   gr$_{}^{-}$  & $\longrightarrow$  &
Y           & + &   Z & + &  gr \\
X$_{}^{+}$  & + &   e$_{}^{-}$   & $\longrightarrow$  &
Y           & + &   Z &   &     \\
\end{tabular}
\end{center}
% =================================================================
Quite a few of the dissociative recombination reactions in the UMIST 
database are branch reactions. We did not use the same branching ratio 
for the corresponding gas--grain collision. Instead, we assumed that 
each branch is equally weighted and that the weights are normalised, 
i.e. $\sum_i g_i = 1$. \\
\indent
We note that gas--grain models have been considered by Willacy et al.
(1998), Willacy \& Langer (2000) and Markwick et al. (2002). These
authors, however, did not include charged grains in their models.
The models computed by Sano et al. (2002) did include charged grains.\\
\indent
We have computed a number of models using gas--phase chemistry only, 
and gas--phase plus mantle and grain chemistry. We now describe the 
implementation of these different models.
% =================================================================
% Table: grain chemistry
% =================================================================
\begin{table}[t]
\caption{List of generalised reactions determining the 
grain chemistry. X$_{}^{+}$ denotes those components 
of the species set which have a neutral counterpart X.}
\begin{center}
\begin{tabular}{rlclclcl}\hline\hline
1. & X$_{}^{+}$               & + &   gr$_{}^{2-}$  
   & $\longrightarrow$  
   & gr$_{}^{-}$              & + &   X           \\
2. & X$_{}^{+}$               & + &   gr$_{}^{-}$ 
   & $\longrightarrow$  
   & gr                       & + &   X           \\     
3. & e$_{}^{-}$               & + &   gr$_{}^{-}$  
   & $\longrightarrow$  
   & gr$_{}^{2-}$             &   &               \\ 
4. & e$_{}^{-}$               & + &   gr  
   & $\longrightarrow$  
   & gr$_{}^{-}$              &   &               \\
5. & e$_{}^{-}$               & + &   gr$_{}^{+}$  
   & $\longrightarrow$  
   & gr                       &   &               \\   
6. & e$_{}^{-}$               & + &   gr$_{}^{2+}$  
   & $\longrightarrow$  
   & gr$_{}^{+}$              &   &               \\
7. & gr$_{}^{+}$              & + &   gr${}^{-}$  
   & $\longrightarrow$  
   & gr                       & + &   gr          \\
8. & gr$_{}^{2+}$             & + &   gr$_{}^{2-}$  
   & $\longrightarrow$  
   & gr                       & + &   gr          \\
9. & gr$_{}^{+}$              & + &   gr$_{}^{2-}$  
   & $\longrightarrow$  
   & gr$_{}^{-}$              & + &   gr          \\
10.& gr$_{}^{-}$              & + &   gr$_{}^{2+}$  
   & $\longrightarrow$  
   & gr$_{}^{+}$              & + &   gr          \\  \hline 
\end{tabular}
\end{center}
\label{list_grain_chemistry}
\end{table}%
% =================================================================

%%%%%%%%%%%%%%%%%%%%%%%%%%%%%%%%%%%%%%%%%%%%%%%%%%%%%%%%%%%%%%%%%%%%
\subsubsection{Model setup} 
The kinetic models we have constructed and applied are listed in 
table~\ref{list_models}. Note that for the fractional abundance of 
grains $x_{\mathrm{gr}^{}} \rightarrow 0$, \texttt{model4} and 
\texttt{model7} reduce to \texttt{model1} and \texttt{model3}, 
respectively.\\
\indent
We considered three different kinetic models of the gas--phase 
chemistry:\\
\noindent  
1) \underline{\texttt{model1}}: The first and the simplest one is the 
reaction scheme introduced by Oppenheimer \& Dalgarno (1974). Species 
M can be represented by any neutral metal atom, and species m 
represents molecular hydrogen. The model is described in detail in
subsection 3.1.1. \\ 
\noindent  
2) \underline{\texttt{model2}}: This kinetic model used two elements, 
similar to the Oppenheimer \& Dalgarno model. All reactions involving 
hydrogen and magnesium only were extracted from the {\sc UMIST} database. 
Note that the reactions involved here differ from the Oppenheimer \& 
Dalgarno model. \\
\noindent  
3) \underline{\texttt{model3}}: All species listed in table 
\ref{list_species_set} were included, giving the most complex kinetic 
model we have studied so far. The model includes the elements: H, He, 
C, O, N, S, Si, Mg, and Fe. The same set of gaseous species was adopted 
by Willacy et al. (1998), Willacy \& Langer (2000) and Markwick et al. 
(2002) in their studies of disk chemistry.\\[.5em]
\noindent  
In addition, we considered four different kinetic models which combine 
gas--phase chemistry with gas--grain processes:\\
\noindent  
1) \underline{\texttt{model4}}: The simplest one is a modification 
of Oppenheimer \& Dalgarno's model (\texttt{model1}) obtained by adding 
the grain and mantle chemistry. The elements are (molecular) hydrogen 
and magnesium, which need to be specified to define the evaporation 
temperature from grain surfaces.\\
\noindent  
2) \underline{\texttt{model5}}: This refers to the kinetic model by 
Umebayashi \& Nakano (1990) as applied by Sano et al. (2000). It is 
described in subsection 3.1.2 .\\
\noindent  
3) \underline{\texttt{model6}}:  This model was designed to improve 
the kinetic model of Sano et al. (2000) by including all species 
listed in table~\ref{list_species_set} containing the 5 elements (H, 
He, C, O, and Mg). We extracted 79 gaseous species from 
table~\ref{list_species_set} which include free electrons. 31 of the 
78 gaseous species have an adsorbed neutral counterpart on the grains, 
so making up the mantle species. Evolving the 79 gaseous species 
requires the extraction of 799 gas--phase reactions from the UMIST 
database. Applying our schemes for grain and mantle chemistry (see 
tables~\ref{list_mantle_chemistry} and \ref{list_grain_chemistry}) 
291 additional reactions were introduced.\\
\noindent  
4) \underline{\texttt{model7}}: Including all the species listed in 
table~\ref{list_species_set}, we construct the most complex kinetic 
model of gas--phase, grain, and mantle chemistry we have studied so 
far. Following the same procedure which was used to set up 
\texttt{model6}, we have evolved 174 gaseous species (including free 
electrons) and 69 mantle species. The same species set (gaseous + 
mantle species) was used by Willacy et al. (1998), Willacy \& Langer 
(2000) and Markwick et al. (2002). 1965 reactions were extracted from 
the {\sc UMIST} database accompanied by 621 additional reactions 
involving grains.
% =================================================================
% Table: kinetic models
% =================================================================
%\begin{table}[htdp]
\begin{table}[t]
\caption{List of kinetic models. Models with indices 1 and 5 refer 
to both prototypes discussed in the text. The remaining models 
refer to the extended models. While models with indices lower 
than 4 refer to pure gas--phase chemistry only, models with indices 
4, 6, and 7 include grain and mantle chemistry.}
\begin{center}
\begin{tabular}{lcccc} \hline\hline
 \hspace*{1.0cm} \hfill & 
 \# elements &  
 \# species  &
 \# reactions &
 grains \\ \hline
\texttt{model1} & 2 &   \ \ \ 5  &   \ \ \ \ \  4     & \\
\texttt{model2} & 2 &   \ \ \ 8  &   \ \ \  18     & \\
\texttt{model3} & 9 & 174        & 1966     & \\
\texttt{model4} & 2 &   \ \  12  &   \ \ \ 24     & $\surd$ \\
\texttt{model5} & 5 &   \ \  13  &   \ \ \ 30     & $\surd$ \\
\texttt{model6} & 5 & 115        &   \ 1090 & $\surd$ \\
\texttt{model7} & 9 & 248        & 2586     & $\surd$ \\ \hline
\end{tabular}
\end{center}
\label{list_models}
\end{table}%
% =================================================================

%%%%%%%%%%%%%%%%%%%%%%%%%%%%%%%%%%%%%%%%%%%%%%%%%%%%%%%%%%%%%%%%%%%%
\subsection{Rate coefficients}
Expressions for the rate coefficients can be obtained from 
macroscopic considerations. The mean collision rate coefficient 
is given by $<\sigma_{\mathrm{c}}{\bf v}>$ where 
$\sigma_{\mathrm{c}}$ and ${\bf v}$ denote the cross section and 
the velocity of the molecules. While the rate coefficients of 
the gas--phase reactions can be calculated easily using information 
in the UMIST database (Le Teuff et al. 2000), computation of the 
rate coefficients for gas--grain processes requires more detailed 
information. We discuss this further in the following sections, but 
note here that we assume a grain radius of $r_{\mathrm{gr}}=10^{-5}$ 
cm in all calculation described in this paper.\\
\noindent
Compared with previous versions of the UMIST database, its latest 
version (Le Teuff et al. 2000) accounts for the temperature dependence 
and specifies the valid temperature range for all reactions. Where 
explicit information about the temperature range is not available, 
the rate coefficients have been attributed a temperature range of 
10 -- 300 K. Since the parameterisation relies on a mathematical fit 
to complex data over a restricted temperature range, the validity of 
the rate coefficients must be checked when used outside their 
temperature range. We used conservative estimates when calculating 
the (temperature-dependent) rate coefficients: if the rate coefficient 
decreases with temperature $T$, we always used the low temperature 
limit $T_1$ as a cut--off value; for lower temperatures the rate 
coefficient was approximated by $T \equiv T_1$. If the rate 
coefficient increases with temperature, we always used the high 
temperature limit T$_2$ as a cut-off value; for higher temperatures 
the rate coeffficient was approximated by $T \equiv T_2$.

%%%%%%%%%%%%%%%%%%%%%%%%%%%%%%%%%%%%%%%%%%%%%%%%%%%%%%%%%%%%%%%%%%%%
\subsubsection{Collisions between ions/electrons and grains}
The rate coefficients were estimated following the paper by Draine 
\& Sutin (1987). Compared to the approximation 
% =================================================================
% figure 4: Draine
% =================================================================
\begin{figure}[t]
\includegraphics[width=.45\textwidth]{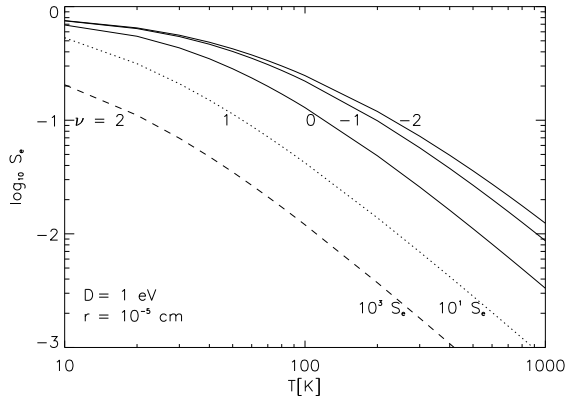}
\caption{Sticking probability of an electron onto neutral grains 
and charged grains as a function of the gas--temperature $T$. 
Curves are labeled by values of $\nu = Ze/q$.\ $D$ and $r$ denotes 
the polarization potential depth $D$ of the grain and the grain 
radius, respectively, while $Ze$ and $q$ refer to the grain charge 
and the charge of the gas--phase particle.}
\label{figure_nishi_1}
\end{figure}
% =================================================================
of Umebayashi \& Nakano (1980) which is used widely in astrochemistry, 
Draine \& Sutin included contributions from the Coulomb potential and 
the polarization of the grain particles induced by the Coulomb field.\\ 
\indent
After some algebra a dimensionless reduced rate coefficient 
$\tilde{J}$ can be extracted from the general expression for 
the rate coefficient. Depending on the ratio $\nu$ of the charge 
$Ze$ of the grain particle to the charge $q$ of the gas--phase 
particle, which relates to the strength of attractive and repulsive 
Coulomb interactions, an approximate formula for the 
reduced rate coefficient can be derived. We applied the equations 
given by Draine \& Sutin (1987), especially equation~(3.3) for 
$\nu = 0$, equation~(3.4) for $\nu < 0$, and equation~(3.5) for 
$\nu > 0$.\\ 
\indent
When calculating the rate of gas--grain particle interaction, we 
took account of the fact that not every collision leads to sticking 
to the grain surface by introducing an average probability $S$ -- 
the ``sticking coefficient". We calculated the sticking coefficients -- 
which depend on the energy, the mass, the charge of the incident 
gas--phase particle, the grain charge and the radius of the grain 
particle -- by applying equations (B5), (B6), (B13) which are given in 
Nishi et al. (1991). While the electron sticking coefficient 
$S_{\mathrm{e^{-}}}$ depends strongly on the temperature and strength 
of the Coulomb interaction (see figure~\ref{figure_nishi_1}), the 
sticking coefficients $S_{\mathrm{X^{+}}}$ for ions depend only weakly 
on the temperature. For simplicity, we assumed $S_{\mathrm{X^{+}}} = 1$ 
which is a good approximation.  

%%%%%%%%%%%%%%%%%%%%%%%%%%%%%%%%%%%%%%%%%%%%%%%%%%%%%%%%%%%%%%%%%%%%
\subsubsection{Collisions between neutral gas--phase particles 
and grain particles}
We have assumed (see reactions listed in 
table~\ref{list_mantle_chemistry}) that collisions between neutral 
% =================================================================
% figure 5: Hollenbach
% =================================================================
\begin{figure}[t]
\includegraphics[width=.45\textwidth]{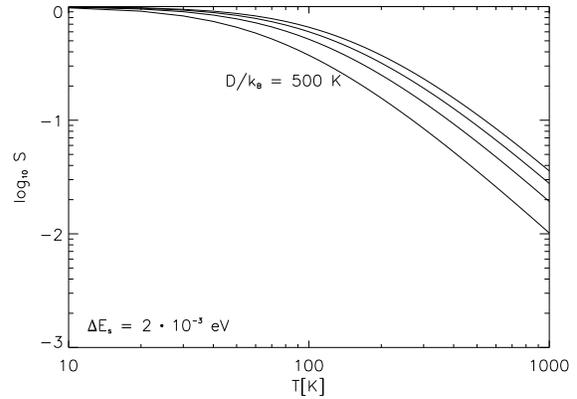}
\caption{Sticking probability of neutral species onto grains as 
a function of the gas--temperature $T$. $\Delta E_{\mathrm{s}}$ 
denotes the amount of energy transferred to the grain particle 
while $D$ is the dissociation energy. The other values of 
$D/k_{\mathrm{B}}$ are 1000, 1500, and 5000 K.}
\label{figure_sticking_neutral}
\end{figure}
% =================================================================
gas--phase particles and grain particles form mantle species. 
In order to specify the rate coefficients for these collisions, 
neutral gas--phase particles are considered to stick onto grain 
surfaces due to thermal adsorption only. The rate coefficient 
can be obtained by averaging the kinetic energy $E_i$ of the 
incident gas--phase particles over the thermal energy distribution, 
assuming that the cross section $\sigma_{\mathrm{c}}$ is 
independent of $E_i$ and is given by the geometrical cross section 
only. The rate coefficient due to thermal adsorption is then 
$k = \sigma_c <v>$ where $<v>$ denotes the mean thermal velocity.\\ 
\indent
When calculating the sticking rate we accounted for the fact that 
more than one collision is needed to trap the gas--phase particle 
onto the grain surface. The procedure for calculating the sticking 
probability $S_{\mathrm{X}}$ of neutral gas--phase species is 
similar to that for calculating $S_{\mathrm{e^{-}}}$. The 
probability $P_{\epsilon}$ that a particle with kinetic energy 
$E_i$ will be adsorbed is approximated by 
$P_{\epsilon} = \exp \{ - \epsilon^2/2 \}$, where $\epsilon$ is 
given by $\epsilon = E_i/\sqrt{D\Delta E_{\mathrm{s}}}$ 
(see Hollenbach \& Salpeter 1970). $D$ and  $\Delta E_{\mathrm{s}}$ 
denote the dissociation energy and the amount of energy 
transferred to the grain particle due to lattice vibration, 
respectively. For each neutral gas--phase component we 
approximated $D$ by its binding energy for physical adsorption 
$E_D$ (see table~\ref{list_evaporation_temperatures}); the value 
of $\Delta E_{\mathrm{s}}$ was approximated by 
$2.0 \times 10^{-3}$ eV.\\
\noindent
Figure~\ref{figure_sticking_neutral} serves as an illustration 
for the temperature dependency of the sticking probability 
$S_{\mathrm{X}}$ for different values of dissociation energy.

%%%%%%%%%%%%%%%%%%%%%%%%%%%%%%%%%%%%%%%%%%%%%%%%%%%%%%%%%%%%%%%%%%%%
\subsubsection{Collisions between grain particles}
The rate coefficients for the four different types of grain--grain 
collisions (see table~\ref{list_grain_chemistry}) were calculated 
by applying equation~(3) of Umebayashi \& Nakano (1990). Here the 
value of the mass density of the grain material $\rho_\mathrm{gr}$ 
is taken to be $\rho_\mathrm{gr} = 3 \ \mathrm{g \ cm^{-3}}$.

%%%%%%%%%%%%%%%%%%%%%%%%%%%%%%%%%%%%%%%%%%%%%%%%%%%%%%%%%%%%%%%%%%%%
\subsubsection{Desorption processes}
Desorption processes cause the ejection of mantle species from 
the grain particles. Here, we considered the thermal desorption 
of mantle species only. The average time for a mantle species 
to overcome the surface binding is given by equation~(2) in 
Hasegawa et al. (1992), replacing the potential energy barrier 
between adjacent surface potential energy wells, $E_{\mathrm{b}}$, 
by the binding energy for physical adsorption $E_D$. The binding 
energy for physical adsorption $E_D$ for each considered mantle 
species is listed in table~\ref{list_evaporation_temperatures}. 
The standard value of $1.5 \times 10^{15} \ \mathrm{cm}^2$ for the 
surface density of sites is assumed.

%%%%%%%%%%%%%%%%%%%%%%%%%%%%%%%%%%%%%%%%%%%%%%%%%%%%%%%%%%%%%%%%%%%%
\subsection{Initial abundances}
While the composition of the gas may vary depending on the 
specific kinetic model, the total particle density $n$ of the 
gas phase can be considered as a conserved quantity. The 
variation in the mean molecular weight $\mu$ of the gas due to 
chemical reactions is negligible as the abundances of molecular 
hydrogen and helium are approximately constant. Hence the value 
$\mu = 2.33$ is kept constant (assuming a solar composition of 
the most abundant elements H and He), and the total particle 
density of the gas phase can be approximated 
by $n = \rho / (\mu m_{\mathrm{H}})$. Here $\rho$ and 
$m_{\mathrm{H}}$ denote the mass density of the gas and the mass 
of the hydrogen nuclei, respectively.\\
\indent
We have adopted the same elemental abundances used by previous 
authors in their studies of the ionisation fraction
(Sano et al. 2000, Fromang et al. 2002, Semenov et al. 2004). 
The abundances of those elements which are not covered by the 
model of Sano et al. (2000) are taken from Semenov et al. (2004). 
Hydrogen and helium excepted, all elements are assumed to be 
depleted in the gas phase compared with solar abundances (shown in 
table~\ref{list_element_abundances}). 
% =================================================================
% Table: Elemental abundances
% =================================================================
\begin{table}[t]%b
\caption{Elemental abundances per hydrogen nucleus.}
\begin{center}
\begin{tabular}{lc} \hline\hline
Element  & Abundances \\ \hline
H      & \hspace*{-1.2cm} $1.00$ \\
He     &  $9.75 \times 10^{-2}$ \\
C      &  $3.26 \times 10^{-4}$ \\
N      &  $1.12 \times 10^{-4}$ \\
O      &  $8.53 \times 10^{-4} $ \\
%Mg     & \hspace*{-.55cm} $1.00 \times 10^{-13}$ \\
Mg     & variable \\
Si     &  $3.58 \times 10^{-5}$ \\
S      &  $1.85 \times 10^{-5}$ \\
%Fe     & \hspace*{-.65cm} $2.74 \times 10^{-9}$ \\ \hline
Fe     & variable \\
gr     & variable \\ \hline
\end{tabular}
\end{center}
\label{list_element_abundances}
\end{table}%
% =================================================================
The fraction $\delta$ of the elements in the gas phase is: 1 
for H and He, 0.2 for C and O, $2.2 \times 10^{-1}$ for N, 
$2.7 \times 10^{-4}$ for Si, and $4.9 \times 10^{-3}$ for S. 
While the reservoir of all the other elements is fixed by these 
values, the abundances of metals (Mg and Fe) are taken as free 
parameters and will be specified elsewhere.\\
\indent
The correct choice of initial abundances of chemical species is 
uncertain as we have no detailed knowledge of how material in
molecular clouds evolves as it collapses to form a protostellar 
disk, or how it evolves during the early stages of disk formation 
and evolution. We used the simplest approach for tackling this 
issue. With exception of hydrogen (which is completely locked in 
molecular form), the elements are taken to be atomic. The 
number of hydrogen nuclei $n_{\mathrm{H}}$ (which is needed to 
get the species' particle concentrations) is approximated by 
$n_{\mathrm{H}} = n / \sum_{i} c_i$, where $c_i$ refers to 
the elemental abundances (per hydrogen nucleus) of element $i$ 
in the gas phase. Note the difference in the notation: 
$x[\mathrm{X}_i]$ denotes the particle concentration of species 
$\mathrm{X}_i$ normalised to the 
total particle density $n$. $x_i$ 
denotes the elemental particle concentration of the 
element $i$.\\
\indent
Since we consider grain particles with different electric charge 
excess, only neutral grains were assumed to exist initially, i.e. 
the total number density $n_{\mathrm{gr}}$ of grain particles is 
given by $n_{\mathrm{gr}} \equiv n[\mathrm{gr}]$ with 
$n[\mathrm{gr}_{}^{\pm}] = n[\mathrm{gr}_{}^{2\pm}] = 0$.

%%%%%%%%%%%%%%%%%%%%%%%%%%%%%%%%%%%%%%%%%%%%%%%%%%%%%%%%%%%%%%%%%%%%
\section{Sources of free electrons}
\label{free_electrons}
In this work we treat the free electrons $\mathrm{e}^{-}$ as an 
independent component of the reaction network and solve a kinetic 
equation for their abundance $x[\mathrm{e}^{-}]$.\\
\indent
Many elementary ionisation processes are considered in the UMIST 
database. Reactions which may cause the formation of free electrons 
are: i) photoreactions ii) reactions due to cosmic--ray particles 
which includes photoreactions induced by cosmic--ray particles and 
iii) chemiionisation reactions.\\
\indent
We have neglected the effects of the stellar and interstellar
ultraviolet radiation field, which drive the photoreactions. There 
is great uncertainty about the amount of unattenuated UV flux that 
reaches the disk surface. In addition the exponential decay of 
this field with increasing column density means that it can only 
act as source of free electrons in a thin skin at the disk 
surface. As we are primarily interested in calculating the 
depth of the magnetically active zones in the disk models, this 
thin surface layer is unimportant.\\
\indent
Galactic cosmic--ray particles are unlikely to reach the 
surface of the inner regions of a protostellar disk due to 
their exclusion by T Tauri winds. Indeed the solar wind 
blocks cosmic--ray particles with energies less than 
$100 \ \mathrm{MeV}$ from entering the inner solar system. 
Preferring a conservative estimate of the electron abundance, 
we did not consider any of the 11 cosmic--ray particle 
reactions listed in the UMIST database. For the same reason, 
we did not consider any of the cosmic--ray--induced 
photoreactions.\\
\indent
Only one chemiionisation reaction, i.e. 
$$
\mathrm{CH} + \mathrm{O} \longrightarrow 
\mathrm{HCO^+} + \mathrm{e^-}
$$
is tabulated in the UMIST database; the importance of this 
particular type of (associative) ionisation for the formation 
of carbon monoxide in interstellar clouds was pointed out by 
Dalgarno et al. (1973).\\[.5em]
\noindent
Other additional sources of free electrons such as the 
contributions due X--ray photons, thermal ionisation, and 
the decay of radioactive elements are analysed in the subsequent 
sections 4.2. 5.1, and 6.0.

%%%%%%%%%%%%%%%%%%%%%%%%%%%%%%%%%%%%%%%%%%%%%%%%%%%%%%%%%%%%%%%%%%%%
\subsection{X--rays}
We have included the effects of X--ray radiation on the chemistry. 
Since the first X--ray observatories were launched, the X--ray 
properties of YSOs have been well documented (see the review by 
Glassgold et al. 2000).\\
\indent
We adopted  values 
$L_{\mathrm{X}} = 10^{30} \ \mathrm{erg \ s}^{-1} $ 
and $k_{\mathrm{B}}T_{\mathrm{X}} = 3 \ \mathrm{keV}$ for 
the total X--ray luminosity $L_{\mathrm{X}}$ and the plasma 
temperature $T_{\mathrm{X}}$, respectively, corresponding 
to a young star of approximately solar mass. Adopting the 
scheme described by Fromang et al. (2002), we modelled the 
X--ray source as a coronal ring of radius 
$10 \ \mathrm{R}_{\odot}$ centered on the star.

%%%%%%%%%%%%%%%%%%%%%%%%%%%%%%%%%%%%%%%%%%%%%%%%%%%%%%%%%%%%%%%%%%%%
\subsection{X--ray ionisation rate}
We assumed that photoelectric absorption of X--ray photons 
dominates while neglecting Compton scattering. For the 
X--ray photon energies considered, this assumption simplifies 
the description of the interaction of X--ray photons with 
matter in a way that depends on the attenuating column 
densities. In addition, the results presented by Igea \& 
Glassgold (1999) indicate that the difference in the total 
X--ray ionisation rates between the 
% =================================================================
% Table: CRP rates
% =================================================================
\begin{table}[t]
\caption{}
\begin{center}
\begin{tabular}{llclclcll}\hline\hline
1. & H$_{2}^{ }$              & $\longrightarrow$  
   & H$_{2}^{+}$              & + &  e$_{}^{-}$  &   & 
   & $0.97 \cdot \zeta_{\mathrm{eff}}$ 
\\
2. & H$_{2}^{}$               & $\longrightarrow$  
   & H$_{}^{+}$               & + &   H          & + &  e$_{}^{-}$
   & $0.03 \cdot \zeta_{\mathrm{eff}}$ 
\\
3. & He                       & $\longrightarrow$  
   & He$_{}^{+}$              & + &   e$_{}^{-}$ &   &
   & $0.84 \cdot \zeta_{\mathrm{eff}}$ \\ \hline
\end{tabular}
\end{center}
\label{table_CRP_rates}
\end{table}%
% =================================================================
% =================================================================
% Figure 6: zeta
% =================================================================
\begin{figure}[t]
\includegraphics[width=.50\textwidth]{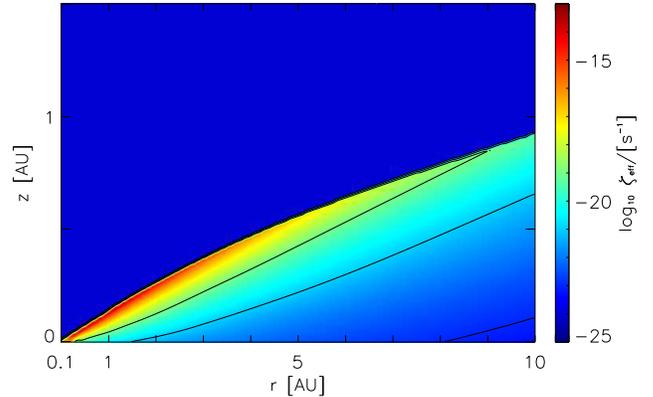}
\caption{The effective X--ray ionisation rate $\zeta_{\mathrm{eff}}$ 
per hydrogen nucleus (including contributions due to thermal 
ionisation of potassium). The disk parameters are $\alpha = 10^{-2}$ 
and $\dot{M} = 10^{-7} \ \mathrm{M_{\odot}yr^{-1}}$. The 
contour lines refer to values of $\zeta_{\mathrm{eff}}$: 
$10^{-19}$, $10^{-21}$, and $10^{-23} \ \mathrm{s}^{-1}$.}
\label{figure_zeta}
\end{figure}
% =================================================================
two cases ``are about the same at small and large values of 
the vertical column density and differ by about one order 
of magnitude at the scattering shoulder".\\
\indent
Following Krolik \& Kallman (1983), we modelled the X--ray flux of 
the optically thin bremsstrahlung spectrum by the exponential law 
$F_{\epsilon} = F_0\exp \{ - \epsilon/k_{\mathrm{B}} 
T_{\mathrm{X}} \}$. For larger X--ray optical depths\footnote{The 
X--ray optical depth is given by $\tau_{\mathrm{X}} = \sigma(E)N$. 
$\sigma$ is the total gas X--ray photoabsorption cross section per 
hydrogen nucleus while $N$ denotes the (particle) column density 
along the line of sight toward the X--ray source.} $\tau_{\mathrm{X}}$, 
the local X--ray flux $F$ per unit energy is attenuated by 
$F = F_{\epsilon} \exp \{ - \tau_{\mathrm{X}} \}$. We further assumed 
that the X--ray ionisation only depends on the elemental composition 
of the gas and is independent of the species composition. Hence, the 
absorption of X--ray photons can be simplified by introducing an 
effective photoionisation cross section (per hydrogen nucleus) 
$\sigma_{\mathrm{eff}} = \sum_{i}^{} x_i^{} \sigma_i$. Here
$\sigma_i$ refers to the partial photoionisation cross section 
of each atomic species and $x_i$ denotes the particle concentration 
of the element $i$. We used the expression for the effective (total) 
photoionisation cross section given by Igea \& Glassgold (1999) for 
the gas--phase depletion of the heavy elements compared to solar 
abundances.\\
\indent
We calculated the effective X--ray ionisation rate 
$\zeta_{\mathrm{eff}}$ for hydrogen by assuming that 
$\zeta_{\mathrm{eff}}$ is dominated by secondary ionisation. 
The number of secondary ionisations $N_{\mathrm{sec}}$ per unit 
of primary photoelectron energy is considered to be independent 
of energy. Finally, the effective X--ray ionisation rate (per hydrogen 
nucleus) $\zeta_{\mathrm{eff}}$ is approximated by 
% =================================================================
% formula: ionisation rate
% =================================================================
\begin{equation}
\zeta_{\mathrm{eff}}^{} \approx 
N_{\mathrm{sec}}^{} \int_{E_{\mathrm{min}}}^{\infty} 
F(E) \sigma_{\mathrm{eff}}^{}dE \ .  
\label{eqn_ionisation_rate}
\end{equation}
% =================================================================
The value of $N_{\mathrm{sec}}$ is about 26 for hydrogen if 
the energy $E$ is measured in units of $\mathrm{keV}$. 
$E_{\mathrm{min}}$ denotes the minimum X--ray photon energy. 
Hence, for values of photon energies less then $E_{\mathrm{min}}$ 
the photon will be absorbed before the disk surface is reached. 
Our choice of $E_{\mathrm{min}} = 0.1 \ \mathrm{keV}$ -- which is 
above the required energy for ionisation of $\mathrm{H_2}$ 
($15.4 \ {\mathrm{eV}}$) -- accounts for absorption in the source 
(see Shang et al. 2002).\\
\indent 
The integral can be transformed into a dimensionless integral 
given for example by equation~(4) in Fromang et al. (2002). By 
applying the method of steepest descent, this dimensionless 
integral can be approximated by an analytic asymptotic formula. 
However, we evaluated the integral numerically and checked 
it against the value obtained by the asymptotic 
approximation.\footnote{Note the typo in equation (5) of 
Fromang et al. (2002), with $b$ instead of $-b$.}\\
\indent
We calculated the X--ray optical depth $\tau_{\mathrm{X}}$ 
along the line of sight between the X--ray source and the 
point in question (as did Fromang et al. 2002).\footnote{Though 
this particular matter is not explicitly described in Semenov 
et al. (2004), the tabulated data indicate they calculated 
$\tau_{\mathrm{X}}$ along the line of sight too.}\\
\indent
For a given effective X--ray ionisation rate 
$\zeta_{\mathrm{eff}}$ we must specify the fraction of 
$\mathrm{H_{\ }^{+}}$ and $\mathrm{H_2^{+}}$ ions produced due 
to X--ray ionisation of $\mathrm{H_2^{}}$. We adopted the values 
given by Cravens \& Dalgarno (1978) taken from calculations of 
the efficiencies that cosmic rays with energy of $1 \ \mathrm{MeV}$, 
traversing a neutral gas of molecular hydrogen, produce these 
particular ions (see table~\ref{table_CRP_rates}). Contributions 
from secondary ionisation were included too. Cravens \& Dalgarno 
found that at proton energies of $1 \ \mathrm{MeV}$ for example, 
the total ionisation rate (which is the sum of primary and secondary 
rates) is larger than the primary rate by a factor of 1.44.\\
\indent
Since our kinetic models do not include photoreactions driven 
by incident FUV flux, we have simplified the terminology used in 
certain sections of this paper. Instead of ``effective X--ray 
ionisation rate'' we often use the phrase ``photoionisation''.

%%%%%%%%%%%%%%%%%%%%%%%%%%%%%%%%%%%%%%%%%%%%%%%%%%%%%%%%%%%%%%%%%%%%
\section{Results}
\label{results}
The majority of our calculations used an $\alpha$--disk model
with $\alpha=10^{-2}$ and ${\dot M}=10^{-7}$ M$_{\odot}$yr$^{-1}$. 
We have evolved the disk chemistry using the kinetic models 
described in section~\ref{chemical_models}, with the primary 
purpose of examining the size of the ``active zone" (the meaning of 
this phrase is explained below). One primary aim of this 
work is to compare and understand the differences in the size of 
the ``active" zones obtained with each of these chemical networks, 
under physical conditions that are otherwise the same.
While a number of studies of the ionisation fraction in protostellar disk 
models have appeared in the literature, they have each used different
kinetic and/or disk models, so that direct comparison between them
has not been possible.
\\
\indent
As outlined in the introduction, the electron fraction is of 
great importance in protoplanetary disks as it determines how 
well the gas is coupled to the magnetic field, and hence whether 
MHD turbulence can be sustained by the MRI. The important 
discriminant is the magnetic Reynolds number ${Re}_{\rm m}^{}$, 
which we define as
% =================================================================
% equation: magnetic reynolds number
% =================================================================
\begin{equation}
{Re}_{\rm m}^{} = \frac{H c_s}{\mu_{\rm m}^{}}
\label{reynolds}
\end{equation}
% =================================================================
where $H$ is the disk semi--thickness, $c_s$ is the sound speed, 
and $\mu_{\rm m}^{}$ is the magnetic diffusivity (not to be confused with
the mean molecular weight). Numerical simulations (e.g. 
Fleming, Stone \& Hawley 2000) indicate that a critical value of the magnetic 
% =================================================================
% figure 7: ionisation fraction referring to Re = 100
% =================================================================
\begin{figure}[t]
\includegraphics[width=.50\textwidth]{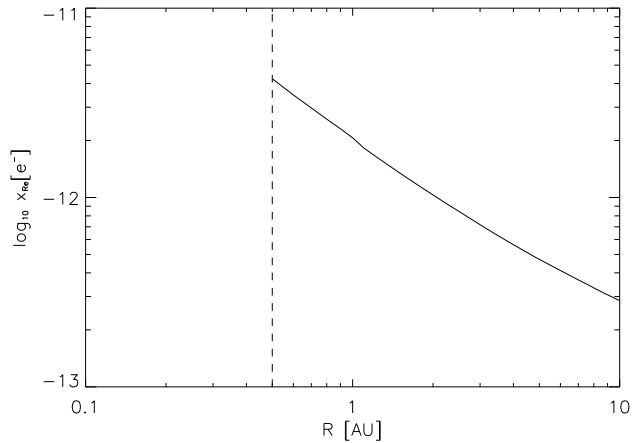}
\caption{- The equilibrium electron concentration
$x_{Re}[\mathrm{e}^{-}]$ for the transition layer separating the active
from the dead zone vs the radial position $R$. A magnetic Reynolds
number of 100 is assumed. Disk regions inside $R \le .5 \ \mathrm{AU}$
are entirely active. The disk parameters are $\alpha = 10^{-2}$ 
and $\dot{M} = 10^{-7} \ \mathrm{M_{\odot}yr^{-1}}$.}
\label{electron-fromang}
\end{figure}
% =================================================================
Reynolds number, ${Re}_{\rm m}^{\rm crit}$, exists such that non linear MHD 
turbulence cannot be sustained if ${Re}_{\rm m}^{}$ falls below 
${Re}_{\rm m}^{\rm crit}$. The value of ${Re}_{\rm m}^{\rm crit}$ depends on 
the field geometry. For a net flux vertical field 
${Re}_{\rm m}^{\rm crit} \simeq 10^2$, and for a zero net flux field 
${Re}_{\rm m}^{\rm crit} \simeq 10^4$. A recent numerical study of MRI driven 
turbulence including the Hall effect indicates that ${Re}_{\rm m}^{\rm crit}$ 
is not particularly sensitive to this effect (Sano \& Stone 2002). The magnetic 
diffusivity $\mu_{\rm m}^{}$ expressed in c.g.s. units is (e.g. Blaes \& Balbus 
1994) 
% =================================================================
% equation: magnetic diffusivity
% =================================================================
\begin{equation}
\mu_{\rm m}^{} = \frac{234}{x[\mathrm{e}_{}^{-}]} T^{1/2}. 
\label{mag_diff}
\end{equation}
% =================================================================
Given the electron fraction $x[\mathrm{e}_{}^{-}]$ from the kinetic 
models, and physical conditions within the disk, we are thus able 
to calculate the spatial variation of ${Re}_{\rm m}^{}$ inside the 
disk models.\\
\indent
In this study we take the value of ${Re}_{\rm m}^{\rm crit} = 100$. 
For values of ${Re}_{\rm m}^{}$ less than this we assume that the disk 
is magnetically decoupled and is ``dead" with respect to sustaining MHD 
turbulence. For ${Re}_{\rm m}^{} \ge 100$ we assume that the disk is 
turbulent. Following previous authors (e.g. Gammie 1996) we use the terms 
``dead" and ``active" zones. We refer to the region where 
${Re}_{\rm m}^{} = 100$ separating the active and dead zones as the 
``transition zone". Values for the ionisation fraction $x[\mathrm{e}_{}^-]$ 
along the transition zone are given in figure~\ref{electron-fromang}, showing 
the values of $x[{\rm e}^{-}]$ required to satisfy ${Re}_{\rm m}^{}=100$ at 
each radial location in the disk.\\
\indent
When presenting our results we typically plot the column density of 
the active zone and of the total disk mass as a function of cylindrical 
radius in the disk, rather than presenting contour plots of the 
free--electron abundance or magnetic Reynolds number. This is to aid 
comparison with results presented by Fromang et al. (2002), who also 
presented their results in this way, and also because it is then clear 
what fraction of the disk mass is active at a particular radial 
location in the disk. \\
\indent
We solved the kinetic equations by integrating over a time interval 
of 100,000 yrs. Hence, the ionisation fraction $x[\mathrm{e}^-]$ is 
a function of time $t$, and in principle so is the location of the 
transition zone. However, for all kinetic models the change in the 
vertical location of the transition zone at all cylindrical radii in 
the computational domain was below the grid resolution 
for $t > 10,000 \ \mathrm{yrs}$. We find that after 100,000 
years, the networks \texttt{model1}, \texttt{model2}, \texttt{model4} 
and \texttt{model5} have reached chemical equilibrium. The more complex 
models, however, (\texttt{model3}, \texttt{model6} and \texttt{model7}), 
do not reach strict chemical equilibrium in this time, although the 
position of the transition zone remains approximately constant. A 
similar situation has been found, for example, by Semenov et al. (2004), 
who also considered complex chemical models. These authors examined the 
evolution for up to $10^6$ years, and found that strict chemical 
equilibrium was still not achieved in optically thick regions.

%%%%%%%%%%%%%%%%%%%%%%%%%%%%%%%%%%%%%%%%%%%%%%%%%%%%%%%%%%%%%%%%%%%%
\subsection{Test case}
\label{test_case_results}
% =================================================================
% figure 8: Fromang
% =================================================================
\begin{figure}[t]
\includegraphics[width=.50\textwidth]{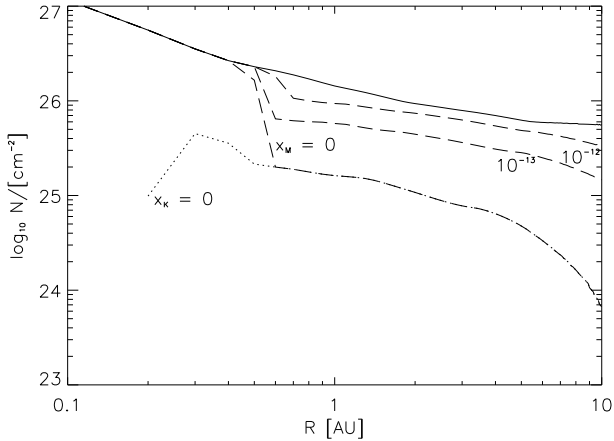}
\caption{Column densities of the whole disk (solid line) and 
of the active zones (dashed and dotted lines) - referring to magnetic 
Reynolds numbers greater than 100 - for values 
$x_{\mathrm{M}^{}} = 0, 10^{-13}, 10^{-12}$. The dotted line refers 
to $x_{\mathrm{K}^{}} = 0$ (i.e. neglecting the contribution due to 
the thermal ionisation of potassium). The disk parameters are 
$\alpha = 10^{-2}$ and 
$\dot{M} = 10^{-7} \ \mathrm{M_{\odot} yr^{-1}}$.}
\label{figure_fromang}
\end{figure}
% =================================================================
In order to test the fidelity of our disk model and X--ray 
ionisation scheme we have recomputed one of the models presented 
in Fromang et al. (2002), who used the kinetic model of Oppenheimer 
\& Dalgarno (1974) to compute the ionisation fraction in 
$\alpha$--disk models. Equation~(\ref{eq_oppenheimer}) shows that 
the equilibrium free--electron abundance $x_{\infty}[\mathrm{e}^{-}]$ 
depends on the effective X--ray ionisation rate 
$\zeta_{\mathrm{eff}}(r,z)$, the number density of the neutral 
particles $n$, and the gas temperature $T$. We calculated the electron 
fraction using the same parameter values used by Fromang et al. (2002), 
and obtained the spatial distribution of ${Re}_m$ using 
equations~(\ref{reynolds}) and (\ref{mag_diff}). We plot the column 
densities of the active zone obtained by our model in 
figure~\ref{figure_fromang}, which should be compared with figures~(4) 
and (5) in Fromang et al. (2002). The agreement is excellent.\\
\indent
For high temperature (and density) regions, we have considered the 
contributions of ground state ionized atoms under the assumption 
of local thermodynamic equilibrium. Despite their low abundances, 
potassium (K) and sodium (Na) provide major contributions the 
ionisation fraction due to the small values of the first ionisation 
potentials, $5.14$ and $4.34 \ \mathrm{eV}$, respectively.\\
\indent
Applying equation~(1) of Fromang et al. (2002), we considered 
the contribution of thermal ionisation of potassium to the electron 
% =================================================================
% figure 9: Oppenheimer & Dalgarno (model1)
% =================================================================
\begin{figure}[t]
\includegraphics[width=.50\textwidth]{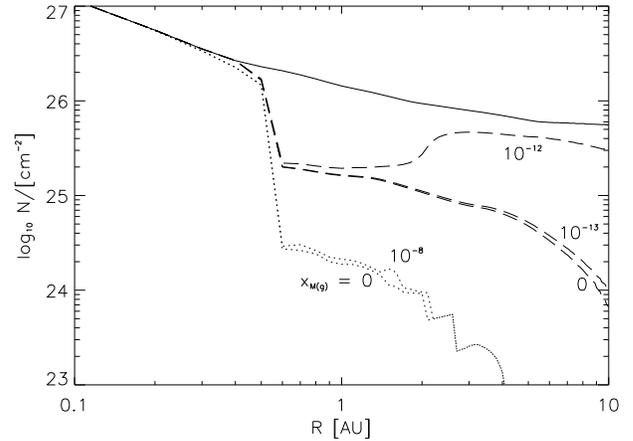}
\caption{- \texttt{model1}/\texttt{model4} - Column densities of 
the whole disk (solid line) and of the active zones (dashed and 
dotted lines) - referring to magnetic Reynolds numbers greater 
than 100 - for different values $x_{\mathrm{M(g)}^{}}$. While the 
dashed lines refer to a value $x_{\mathrm{gr}^{}} = 0$ (gas--phase 
ch.), the dotted lines refer to a value 
$x_{\mathrm{gr}^{}} = 10^{-12}$ (gas--grain ch.). Especially, for a 
grain free environment no dead zones are observed 
above values $x_{\mathrm{M}^{}} \ge 10^{-11}$. The disk parameters 
are $\alpha = 10^{-2}$ and 
$\dot{M} = 10^{-7} \ \mathrm{M_{\odot} yr^{-1}}$.}
\label{column_model1}
\end{figure}
% =================================================================
abundance. Neglecting this particular contribution leads to smaller 
values in the column densities of the active zone in the very inner 
regions only as shown in figure~\ref{figure_fromang}. Including it 
ensures that the whole disk within about 0.4 AU is active.

%%%%%%%%%%%%%%%%%%%%%%%%%%%%%%%%%%%%%%%%%%%%%%%%%%%%%%%%%%%%%%%%%%%%
\subsection{Gas--phase chemistry}%
\label{gas_phase_results}
\noindent
\underline{\bf \texttt{model1}}\\[0.5em]
\noindent
In section~\ref{test_case_results} the electron fraction predicted by the 
Oppenheimer \& Dalgarno (1974) kinetic model was obtained 
using an analytic approximation, giving rise to the results 
in figure~\ref{figure_fromang}. 
We have also solved the full kinetic equations of \texttt{model1}, 
and can compare the exact equilibrium value of $x[{\rm e}^{-}]$ 
with the equilibrium value $x_{\infty}[\mathrm{e}^{-}]$ given by 
equation~(\ref{eq_oppenheimer}). \\
\indent
We find that the solution for $x_{\infty}[\mathrm{e}^{-}]$ obtained 
from equation~(\ref{eq_oppenheimer}) is very accurate if the exact 
equilibrium values for neutral metal atoms, $x_{\mathrm{M}}^{}$, is 
used in equation~(\ref{eq_oppenheimer}). Comparing our results in 
figure~\ref{column_model1} to those presented in 
figure~\ref{figure_fromang} and Fromang et al. (2002), we obtain 
agreement when the metal abundance $x_{\rm M}^{}=0$. We find 
significant differences, however, when the metal abundance is 
$x_{\rm M}^{}=10^{-13}$ or $10^{-12}$. The differences can be traced 
to the assumption, used to obtain figure~\ref{figure_fromang}, that 
the equilibrium concentration of neutral metals, $x_{\infty}[{\rm M}]$, 
is constant and equal to the total fractional abundance of metals. In 
fact the full solution to the kinetic equations predicts that 
$x[{\rm M}]$ can vary by up to two orders of magnitude, especially within 
$R \le 2$ AU, where the effects of the metals in increasing the vertical 
size of the active zone is correspondingly diminished.\\
\indent
As already discussed at some length by Fromang et al. (2002), the 
addition of small quantities of metals (much below solar abundance) 
to the reaction scheme causes the predicted size of the active zone 
to increase. This is because the charge--transfer reaction between 
the metal and molecular ion reduces the concentration of molecular 
ions which recombine with electrons rapidly. The recombination rate 
% =================================================================
% figure 10: column densities of the active zone (model2)
% =================================================================
\begin{figure}[t]
\includegraphics[width=.50\textwidth]{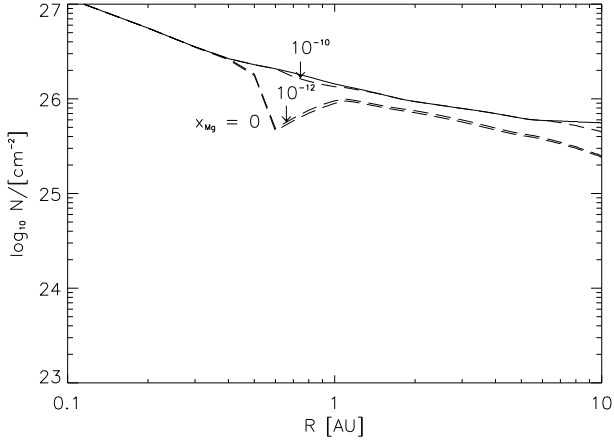}
\caption{- \texttt{model2} - Column densities of the whole disk 
(solid line) and of the active zones (dashed lines) - referring 
to magnetic Reynolds numbers greater than 100 - for values 
$x_{\mathrm{Mg}^{}} = 0, 10^{-12},10^{-10}$. The disk parameters 
are $\alpha = 10^{-2}$ and 
$\dot{M} = 10^{-7} \ \mathrm{M_{\odot} yr^{-1}}$.}
\label{column_model2}
\end{figure}
% =================================================================
between metal ions and electrons is much smaller, leading to a higher 
equilibrium electron fraction. Comparing the rate coefficient for 
molecular ion--electron recombination ($\tilde{\alpha}$) with that 
for metal ion--electron recombination ($\tilde{\gamma}$) illustrates 
this point (see table~\ref{list_oppenheimer}).\\
\indent
Moving out from the centre of the disk, the features observed in the
size of the active zone can be explained as follows. Interior to 
$R \le 0.5$ AU the disk is completely active due to thermal ionisation 
of potassium. For values of $x_{\rm M}^{}=10^{-13}$ and $10^{-12}$ the 
region between $0.5 \le R \le 2$ AU maintains a deep dead zone because 
the metal concentration is too small to prevent H$^+_2$ being the 
dominant ion, whose continued presence maintains a low electron fraction. 
Exterior to 2 AU a metal concentration significantly enlarges the active 
zone. Primarily this is because a lower ionisation degree is required at 
larger radius to obtain an active layer with ${Re}_{\rm m}^{} \ge 100$, as 
shown by figure~\ref{electron-fromang}. A relatively low abundance of metals
is required to achieve this threshold at radii beyond 2 AU. Note that for metal 
concentrations $x_{\rm M}^{} \ge 10^{-11}$ the whole of our disk model is 
active.\\[.5em]
\noindent
\underline{\bf \texttt{model2}}\\[0.5em]
\noindent
This model takes the elements hydrogen and magnesium, and extracts 
all the relevant species and reactions from the UMIST database. 
% =================================================================
% figure 11: column density of the active zone (model3)
% =================================================================
\begin{figure}[t]%[thb]
\includegraphics[width=.50\textwidth]{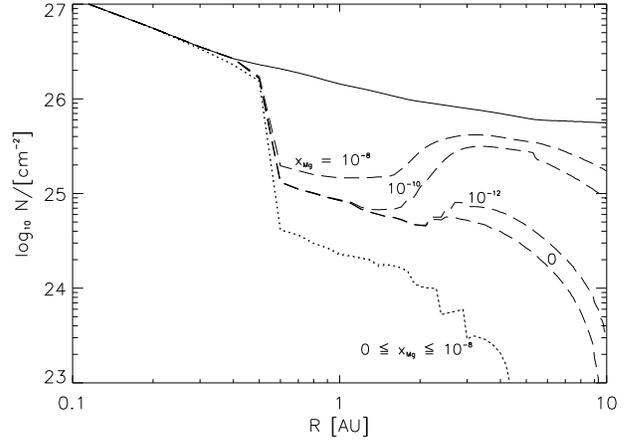}
\caption{- \texttt{model3}/\texttt{model7} - Column densities of 
the whole disk (solid line) and of the active zones (dashed and 
dotted lines) - referring to magnetic Reynolds numbers greater than 
100 - for values $x_{\mathrm{Mg}^{}} = 0,10^{-12},10^{-10},10^{-8}$ 
($x_{\mathrm{Fe}^{}} = 0$). The dashed lines refer to a value 
$x_{\mathrm{gr}^{}} = 0$ (gas--phase ch.), while the dotted 
lines refer to $x_{\mathrm{gr}^{}} = 10^{-12}$ (gas--grain ch.). 
The disk parameters are $\alpha = 10^{-2}$ and 
$\dot{M} = 10^{-7} \ \mathrm{M_{\odot} yr^{-1}}$.}
\label{column_model3}
\end{figure}
% =================================================================
Solving the resulting reaction set for \texttt{model2} produces an
active zone with much larger column density than obtained 
using \texttt{model1}. This is illustrated by comparing 
figures~\ref{column_model1} and \ref{column_model2}. It is 
clear that this effect has nothing to do with the presence of 
metals, as the active zone is large in their absence. Instead 
the hydrogen chemistry contained in the UMIST database leads 
to a dominant molecular ion H$^+$ whose recombination time is 
much longer than that (for H$_2^+$) in the Oppenheimer \& 
Dalgarno model. The active zone predicted by \texttt{model2} 
is also much larger than predicted by a much more complex 
gas--phase chemistry (\texttt{model3}) so we neglect it from 
further discussion as it seems to be unreliable.\\[.5em]
\noindent
\underline{\bf \texttt{model3}}\\[0.5em]
\noindent
This model was constructed by extracting all species and 
reactions from table~\ref{list_species_set} and the UMIST 
database containing the elements H, He, C, O, N, S, Si, Mg, 
Fe. The results obtained when solving the \texttt{model3} 
reaction set are shown in figure~\ref{column_model3}. 
Comparing this with figure~\ref{column_model1}, it is apparent 
that \texttt{model3} predicts smaller active zones than 
\texttt{model1}, both with and without metals. The metal atoms 
included in \texttt{model3} were magnesium and iron, however we 
find essentially no difference in the ionisation structure of 
the disk when iron is neglected.\\
\indent
Apart from the chemiionisation reaction discussed in 
section~\ref{free_electrons}, free electrons can only be 
% =================================================================
% figure 12: column density of the active zone (model6)
% =================================================================
\begin{figure}[t]
\includegraphics[width = .50\textwidth]{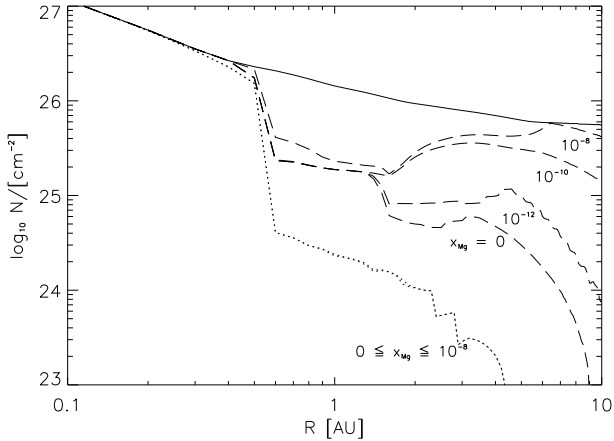}
\caption{- \texttt{model6} - Column densities of the whole 
disk (solid line) and of the active zones (dashed and dotted lines) 
- referring to magnetic Reynolds numbers greater than 100 - 
for values $x_{\mathrm{Mg}^{}} = 0, 10^{-12}, 10^{-10}, 10^{-8}$. 
The dashed lines refer to a value $x_{\mathrm{gr}^{}} = 0$ (gas--phase 
ch.), while the dotted lines refer to $x_{\mathrm{gr}^{}} = 10^{-12}$ 
(gas--grain ch.). The disk parameters are $\alpha = 10^{-2}$ and 
$\dot{M} = 10^{-7} \ \mathrm{M_{\odot} yr^{-1}}$.}
\label{column_model6}
\end{figure}
% =================================================================
created by the photoionisation of $\mathrm{H}_2$ and 
$\mathrm{He}$. The inclusion of $\mathrm{He}$ thus increases 
the ionisation rate compared to that in \texttt{model1}. 
Removal of free electrons occurs primarily through dissociative 
recombination reactions with ions. Most of these ions are 
formed through charge transfer--reactions that are initiated by 
the photoionisation of hydrogen and helium. This leads to the 
formation of about 100 ionized gas--phase species in 
\texttt{model3} which may recombine with electrons. \\
\indent
For $x_{\mathrm{Mg}} = 0$, the only ion in \texttt{model1} 
is $\mathrm{H}_2^+$. In \texttt{model3} the dominant molecular 
ions are $\mathrm{HCNH}_{}^+$, $\mathrm{H_4^{}C_2^{}N_{}^+}$, 
$\mathrm{CNC}_{}^+$ and $\mathrm{NO}_{}^+$. Although their 
concentration values are below the value of $x[\mathrm{H}_2^+]$ 
obtained using \texttt{model1}, they destroy $\mathrm{e}_{}^-$ 
more efficiently than $\mathrm{H}_2^+$.\\
\indent
For $x_{\mathrm{Mg}} \ne 0$, $\mathrm{Mg}_{}^+$ becomes by far 
the most dominant ion in the active zone as $x_{\mathrm{Mg}}$ 
is increased. However, due to the much larger rate coefficients 
for dissociative recombination reactions involving ions such as 
$\mathrm{NH}_4^+$, $\mathrm{NO}_{}^+$ and $\mathrm{HCNH}_{}^+$, 
these more complex species remain the dominant species that 
recombine with free electrons. Hence, $\mathrm{Mg}_{}^+$ ions 
change rather than dominate the destruction rate for free 
electrons. This remains the case even for relatively large metal 
abundances up to $x_{\rm Mg}=10^{-8}$, such that a dead zone 
remains for this value.\\
\indent 
Motivated by the results of Sano et al. (2000) and those 
obtained using \texttt{model3} which show the nitrogen bearing 
species to be important, we studied 
the effect of nitrogen on the ionisation fraction by 
implementing \texttt{model6}, which does not include nitrogen 
bearing species. We considered the gas--phase kinetics of 
\texttt{model6} by setting $x_{\mathrm{gr}^{}} = 0.$ The molecular 
ions which dominate the destruction of $\mathrm{e}_{}^-$ now differ 
from those in \texttt{model3}. Dissociative recombination reactions 
with ions such as $\mathrm{H_3^{}O_{}^+}$, $\mathrm{HCO_{}^+}$ (for 
$x_{\mathrm{Mg}} = 0$) and $\mathrm{Mg_{}^+}$, 
$\mathrm{H_3^{}O_{}^+}$, $\mathrm{C_3^{}H_3^+}$ and 
$\mathrm{CH_3^{}CO_{}^+}$ (for $x_{\mathrm{Mg}} \ne 0$) dominate the 
destruction process of $\mathrm{e}_{}^-$. However, we find that the 
location of the transition zone is only significantly modified in the 
regions interior to $R=2$ AU, as may be observed when comparing 
figures~\ref{column_model3} and \ref{column_model6}. The neglect of 
nitrogen bearing species increases the size of the active zone, 
indicating that these species are important in determining the 
ionisation balance in protostellar disks. However, it is also clear 
that the remaining non--nitrogen bearing molecular ions are equally 
important, and lead to the formation of a smaller active zone than 
predicted by the simple Oppenheimer \& Dalgarno (1974) model.
% =================================================================
% figure 13: column density of the active zone (model5)
% =================================================================
\begin{figure}[t]
\includegraphics[width = .50\textwidth]{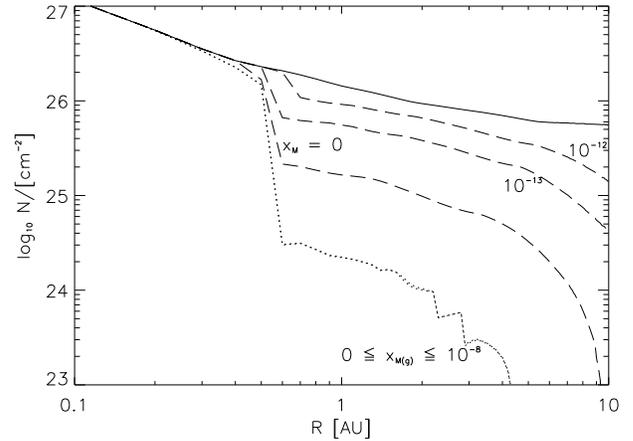}
\caption{- \texttt{model5} - Column densities of the whole 
disk (solid line) and of the active zones (dashed and dotted 
lines) - referring to magnetic Reynolds numbers greater than 100 - 
for different values of $x_{\mathrm{M(g)}^{}}$. While the 
dashed lines refer to $x_{\mathrm{gr}^{}} = 0$ (gas--phase ch.), 
the dotted lines refer to $x_{\mathrm{gr}^{}} = 10^{-12}$ (gas--grain 
ch.). The disk parameters are $\alpha = 10^{-2}$ and 
$\dot{M} = 10^{-7} \ \mathrm{M_{\odot} yr^{-1}}$.}
\label{column_model5}
\end{figure}
% =================================================================
Even a metal abundance of $x_{\rm Mg}^{}=10^{-8}$ is insufficient
to make the disk fully active due to the continued dominance
of the molecular ions.\\
\indent
We conclude that the additional molecular ions generated in more 
complex reaction networks generally lead to a smaller fractional 
abundance of free electrons than the simple Oppenheimer \& Dalgarno 
(1974) reaction scheme. Hence a larger dead zone occurs within 
protostellar disks than is predicted using the simple scheme in 
conjunction with the rate coefficients listed in 
table~\ref{list_oppenheimer}. However, in section~\ref{discussion} 
we show that by making modest changes to the rate coefficients 
in the Oppenheimer \& Dalgarno model good agreement may be obtained 
with the full UMIST database model for certain metal abundances.

%%%%%%%%%%%%%%%%%%%%%%%%%%%%%%%%%%%%%%%%%%%%%%%%%%%%%%%%%%%%%%%%%%%%
\subsection{Gas--grain chemistry}
\label{result-grain}
\begin{figure}[t]
\includegraphics[width = .50\textwidth]{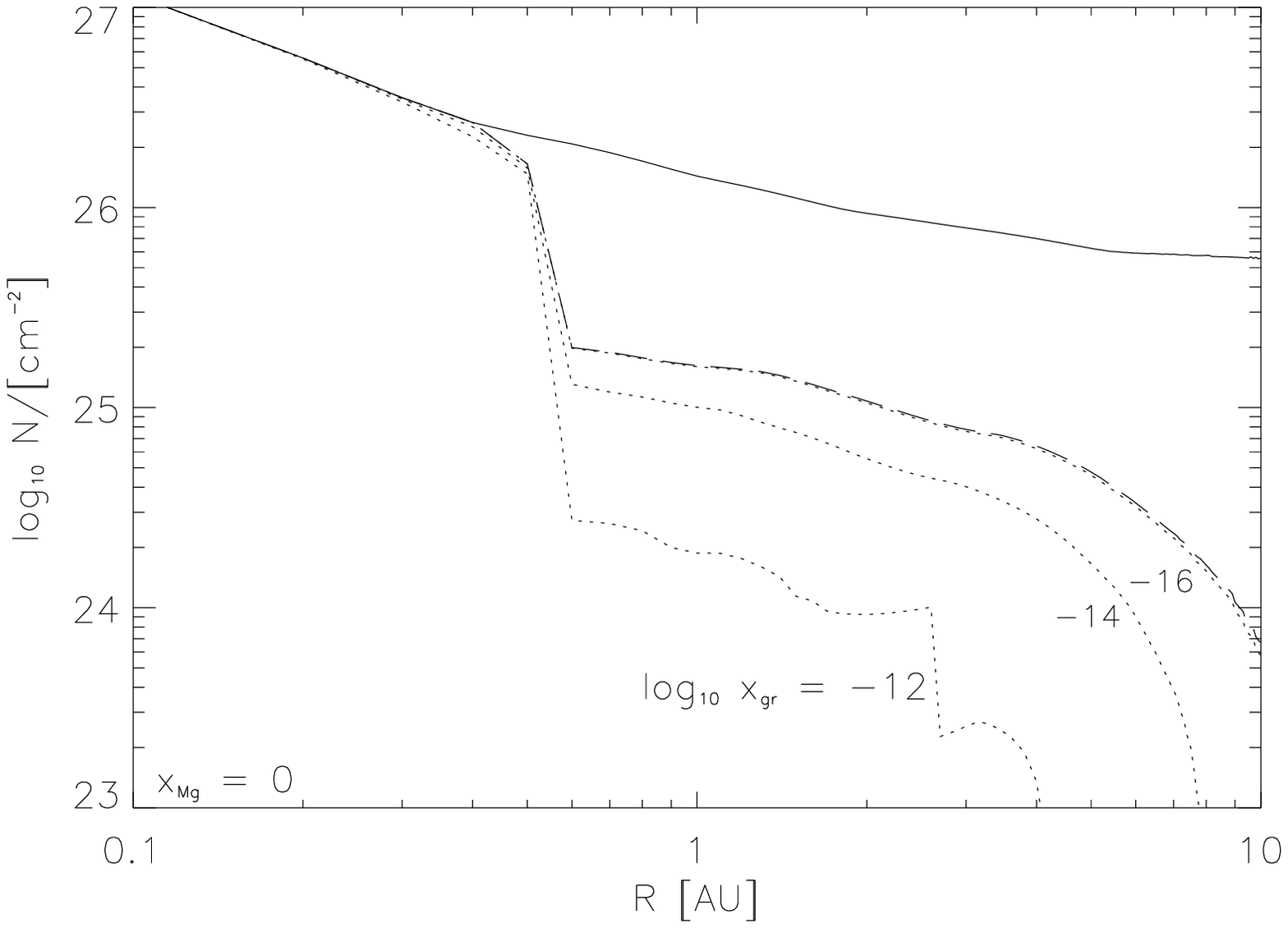}
\caption{- \texttt{model4} - Column densities of the whole disk 
(solid line) and of the active zones (dashed and dotted lines) - 
referring to magnetic Reynolds numbers greater than 100 - for 
different values of $x_{\mathrm{gr}^{}}$ (while 
$x_{\mathrm{Mg}^{}} = 0$ is assumed). While the dashed line refers 
to $x_{\mathrm{gr}^{}} = 0$ (gas--phase ch.), the dotted lines refer 
to $x_{\mathrm{gr}^{}} \ne 0$ (gas--grain ch.). The disk parameters 
are $\alpha = 10^{-2}$ and 
$\dot{M} = 10^{-7} \ \mathrm{M_{\odot} yr^{-1}}$.}
\label{column_model4_c_gr}
\end{figure}

\begin{figure}[t]
\includegraphics[width = .50\textwidth]{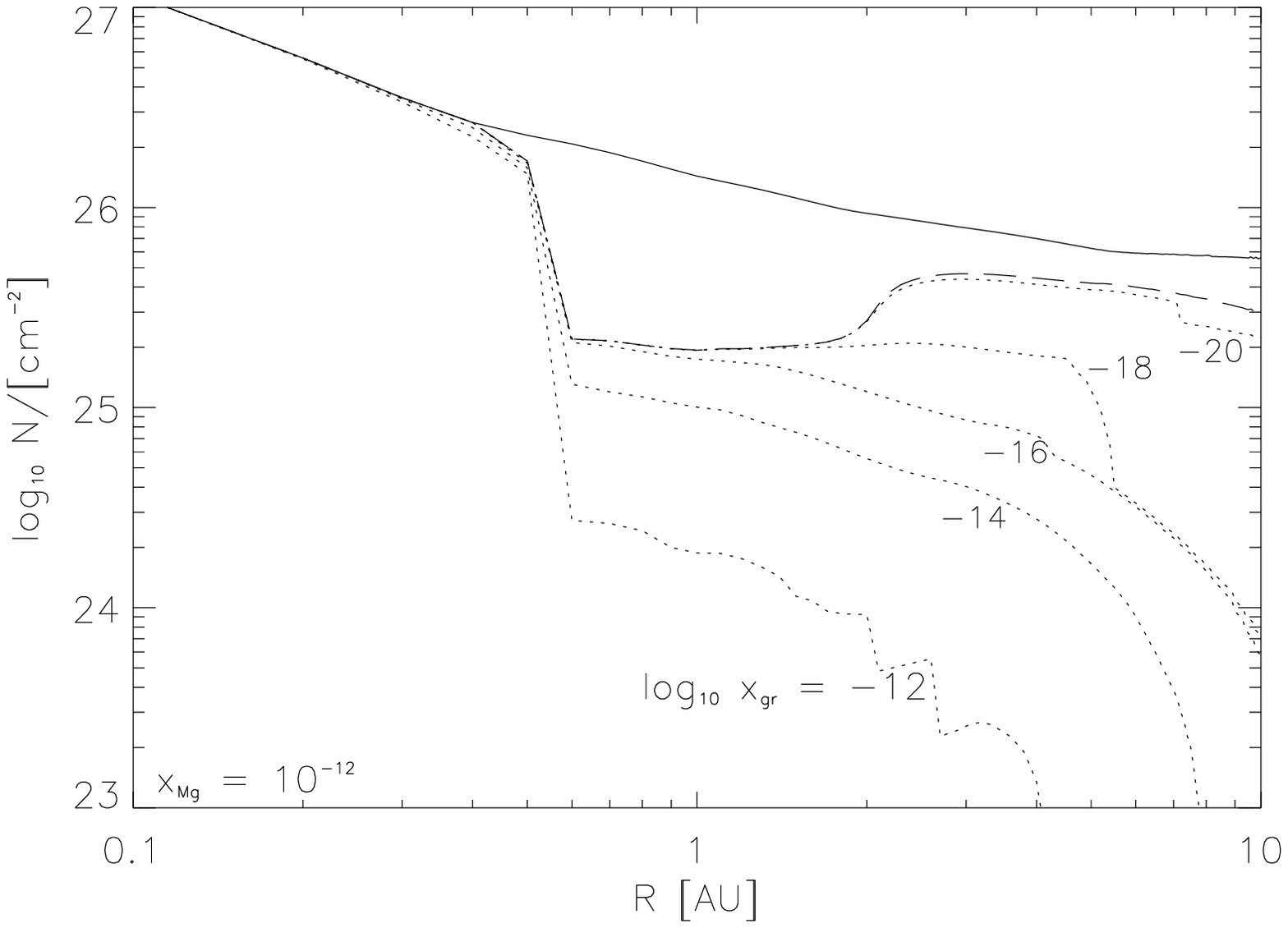}
\caption{- \texttt{model4} - Column densities of the whole disk 
(solid line) and of the active zones (dashed and dotted lines) - 
referring to magnetic Reynolds numbers greater than 100 - for different 
values $x_{\mathrm{gr}^{}}$ ($x_{\mathrm{Mg}^{}} = 10^{-12}$ is assumed). 
While the dashed line refers to a value $x_{\mathrm{gr}^{}} = 0$ 
(gas--phase ch.), the dotted lines refer to $x_{\mathrm{gr}^{}} \ne 0$ 
(gas--grain ch.). The disk parameters are $\alpha = 10^{-2}$ 
and $\dot{M} = 10^{-7} \ \mathrm{M_{\odot} yr^{-1}}$.}
\label{column_model4_c_Mg}
\end{figure}

\begin{figure}[t]
\includegraphics[width=.50\textwidth]{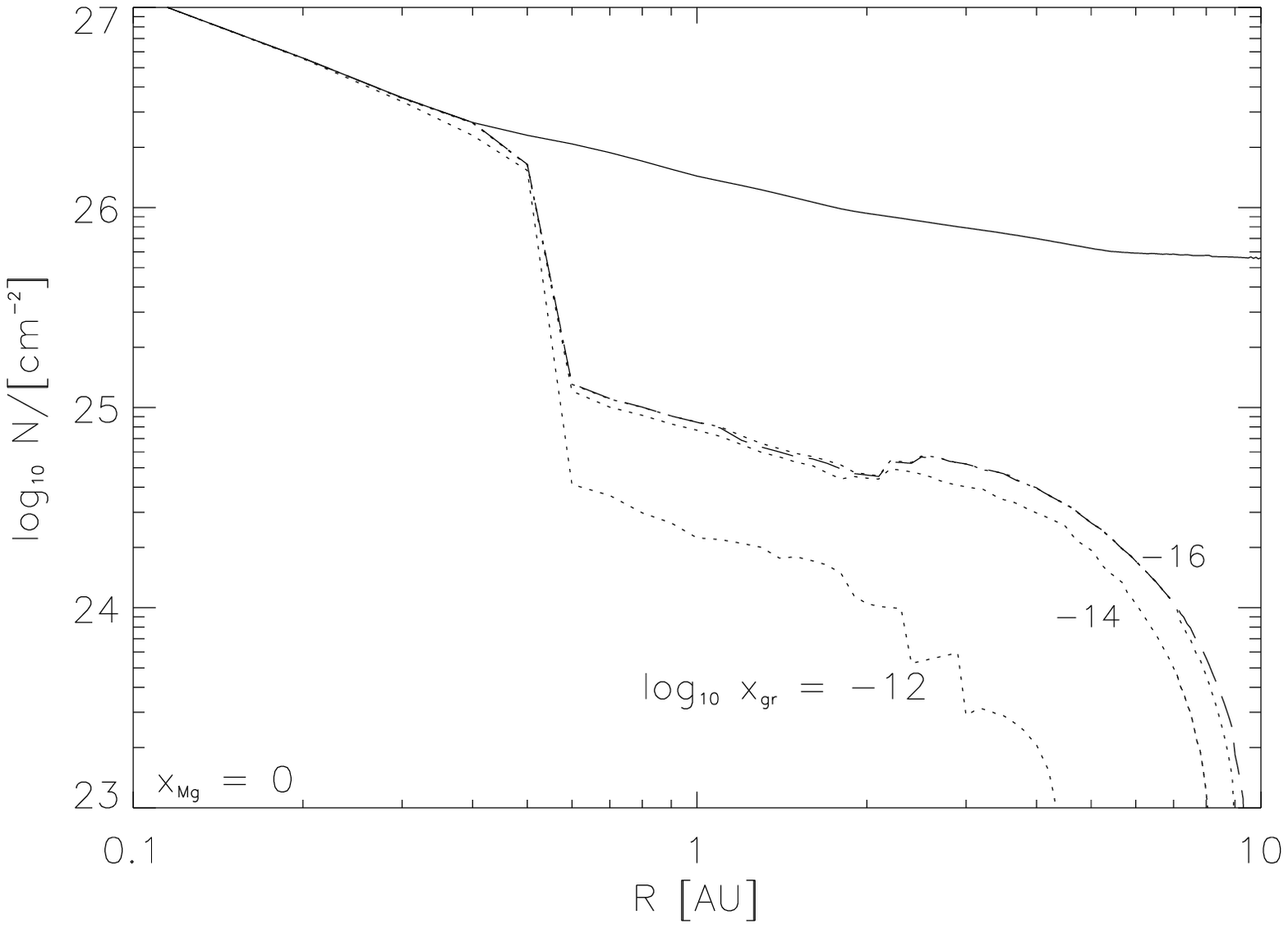}
\caption{- \texttt{model7} - Column densities of the whole disk 
(solid line) and of the active zones (dashed and dotted lines) - 
referring to magnetic Reynolds numbers greater than 100 - for 
different values of $x_{\mathrm{gr}^{}}$ (while 
$x_{\mathrm{Fe}^{}} = x_{\mathrm{Mg}^{}} = 0$ is assumed). While 
the dashed line refers to $x_{\mathrm{gr}^{}} = 0$ (gas--phase ch.), 
the dotted lines refer to $x_{\mathrm{gr}^{}} \ne 0$ (gas--grain ch.). 
The disk parameters are $\alpha = 10^{-2}$ and 
$\dot{M} = 10^{-7} \ \mathrm{M_{\odot} yr^{-1}}$.}
\label{column_model7_c_gr}
\end{figure}

\begin{figure}[t]
\includegraphics[width=.50\textwidth]{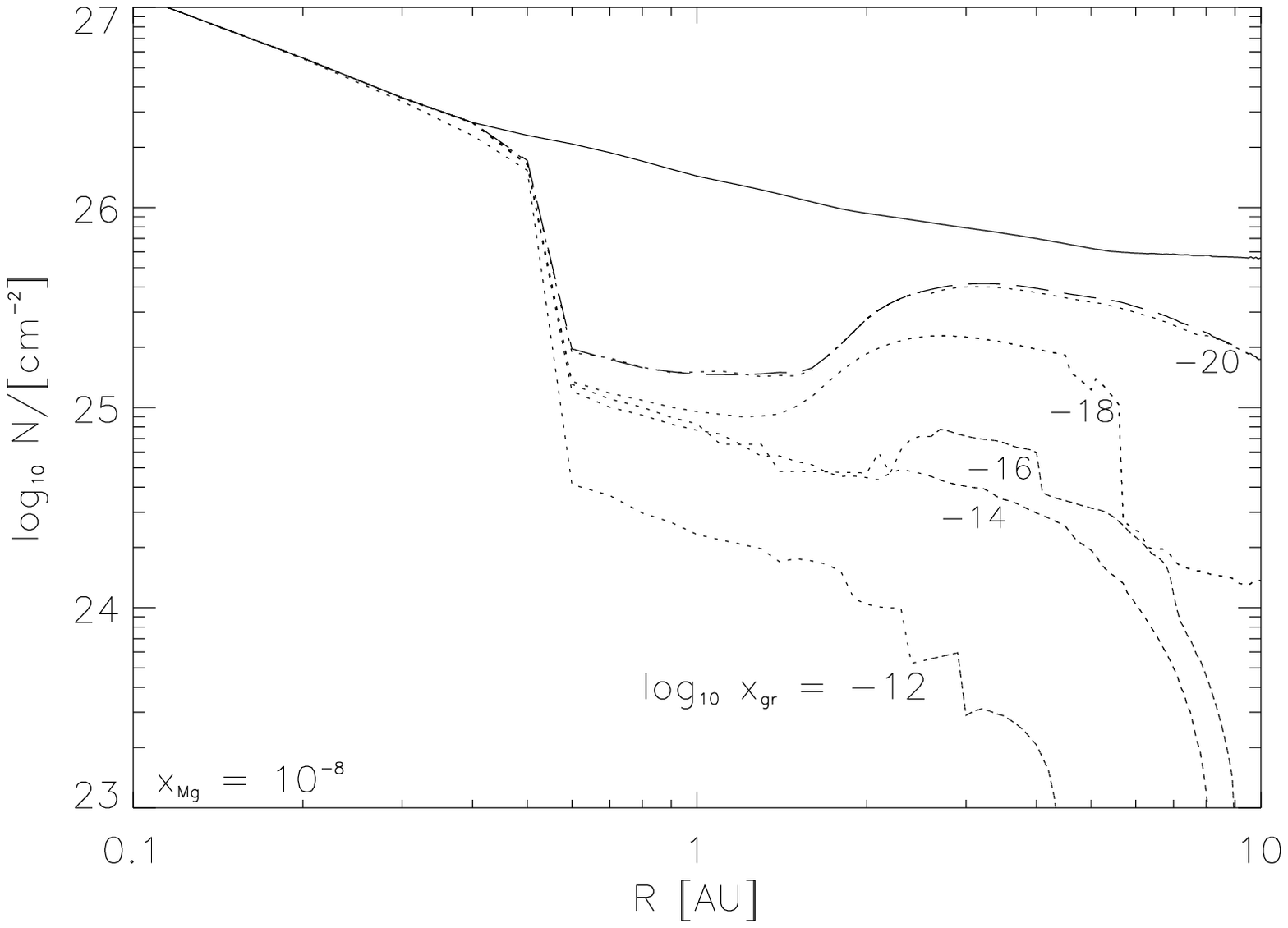}
\caption{- \texttt{model7} - Column densities of the whole disk 
(solid line) and of the active zones (dashed and dotted lines) - 
referring to magnetic Reynolds numbers greater than 100 - for 
different values of $x_{\mathrm{gr}^{}}$ (while 
$x_{\mathrm{Fe}^{}} = 0, x_{\mathrm{Mg}^{}} = 10^{-8}$ is assumed). 
While the dashed line refers to $x_{\mathrm{gr}^{}} = 0$ (gas--phase 
ch.), the dotted lines refer to $x_{\mathrm{gr}^{}} \ne 0$ (gas--grain 
ch.). The disk parameters are $\alpha = 10^{-2}$ and 
$\dot{M} = 10^{-7} \ \mathrm{M_{\odot} yr^{-1}}$.}
\label{column_model7_c_Mg}
\end{figure}

\noindent 
We now present results from kinetic models that include gas--phase 
reactions, ``mantle chemistry", and ``grain chemistry" (see 
section~\ref{extended_models} for a definition of these terms). We 
consider grains of a fixed size ($r_{\mathrm{gr}}=10^{-5}$ cm) and 
with the possibility of having five different charges: $\mathrm{gr}$, 
$\mathrm{gr}_{}^{\pm}$, and $\mathrm{gr}_{}^{2\pm}$. Our motivation 
for examining these models is to first address the question of how 
the various gas--phase models change and differ from one another when 
grains are introduced. Second, to examine the effects of depleting 
the concentration of small grains on the chemistry as a crude means of
modelling the effects of grain growth, and third to define under 
which conditions gas--grain chemistry allows the existence of a 
significant active zone.\\[0.5em]
\noindent
\underline{\bf \texttt{model4}/\texttt{model5}}\\[0.5em]
\noindent
The first reaction scheme including gas--grain chemistry used to 
examine the issue of the ionisation fraction in protostellar disks 
was introduced by Sano et al. (2000). Their model is equivalent to 
our \texttt{model5}. We begin our discussion by first examining the 
gas--phase chemistry produced by \texttt{model5} (i.e. with 
$x_{\mathrm{gr}^{}} = 0$). The results are shown in 
figure~\ref{column_model5}. Comparing figure~\ref{column_model5} 
with figure~\ref{figure_fromang}, we find that \texttt{model5} is in close 
agreement with \texttt{model1} (i.e. the Oppenheimer \& Dalgarno model for 
which $x[\mathrm{e}_{}^{-}]$ is calculated from 
equation~\ref{eq_oppenheimer}). We note that the introduction of grain 
chemistry dramatically decreases the size of the active zone, and gives 
agreement with \texttt{model4}.\\
\indent
We now discuss the results obtained with \texttt{model4}. This model 
is an extension of the Oppenheimer \& Dalgarno (1974) reaction 
scheme (\texttt{model1}) which includes gas--grain interactions. The 
original model included a generalised molecule and a generalised 
metal. As we need to specify the evaporation temperature from the 
grains for these components, we use the species H$_2$ and Mg.\\
\indent
The results obtained with \texttt{model4}, but with magnesium not 
being included, are presented in figure~\ref{column_model4_c_gr} 
(we discuss the results with magnesium included below). In the 
limit of $x_{\mathrm{gr}^{}} = 0$ this model reduces to \texttt{model1}. 
The introduction of grains with concentration 
$x_{\mathrm{gr}^{}} = 10^{-12}$ leads to a dramatic reduction in the 
size of the active zone, as the grains are very efficient at sweeping 
up most of the free electrons. This result is in basic agreement with 
expectations.\\
\indent
We now consider the effect of reducing the grain concentration. 
Figure~\ref{column_model4_c_gr} shows the effect of including grains 
with concentrations $x_{\mathrm{gr}^{}} = 10^{-14}$ and $10^{-16}$. 
A grain concentration of $x_{\mathrm{gr}^{}} = 10^{-16}$ is required 
for the gas--grain chemistry to generate an active zone that is the 
same size as the pure gas--phase chemistry. This result is interesting, 
and at first sight rather puzzling, since figure~\ref{electron-fromang} 
shows that the electron fraction required for the disk to be active is 
between  $x[{\rm e}^-] = 3 \times 10^{-13}$ and $4 \times 10^{-12}$. If 
the grains simply act as a means of sweeping up the free electrons, how 
can a very small concentration of grains such as 
$x_{\mathrm{gr}^{}} = 10^{-14}$ significantly affect the position of the 
transition between active and dead zones?\\
\indent
In the simple \texttt{model4} without magnesium, the grain particles 
are involved in the main destruction paths for $\mathrm{H}_2^+$ and 
$\mathrm{e}_{}^-$. While free electrons are mostly destroyed by 
neutral grains through the reaction 
% =================================================================
$$
\hspace*{-.3cm} \mathrm{e}_{}^{-} + \mathrm{gr} \hspace*{.3cm} 
\longrightarrow \mathrm{gr}_{}^- \ ,
$$
% =================================================================
$\mathrm{H}_2^+$ is destroyed by recombination with singly 
negatively charged grain particles 
% =================================================================
$$
\mathrm{H}_{2}^{+} + \mathrm{gr}_{}^- \longrightarrow \mathrm{H}_2^{} +
\mathrm{gr}.
$$
% =================================================================
These two processes form a loop that recycles neutral 
grain particles, ensuring that they always participate actively 
in the chemical evolution. The equilibrium electron fraction 
$x_\infty[{\rm e^-}]$ is determined by a balance between production 
and destruction processes. Apparently only a small number of active 
neutral grains are required to modify the equilibrium ionisation 
structure in a disk due to the high rate with which electrons accrete 
onto the grains.\\
\indent
In order to reduce the destruction of $\mathrm{e}_{}^{-}$ by neutral 
grains one may limit the neutral grain concentration $x[\mathrm{gr}]$ 
by:\\
\indent
({\it i}). \hspace*{.03cm} reducing $x_{\mathrm{gr}}$ -- the total grain 
concentration \\
\indent
({\it ii}). limit the formation of neutral grains by breaking\\
\indent
\hspace*{0.6cm}
the cycle described above.\\
Satisfying option ({\it i}) requires the value 
$x_{\mathrm{gr}^{}} = 10^{-16}$ already discussed. By contrast, we find 
that the transition from gas--grain dominated chemistry to gas--phase 
chemistry can be achieved for values $x_{\mathrm{gr}^{}} \sim 10^{-14}$ 
if the cycle is short--circuited by setting the rate coefficient of the 
reaction 
$\mathrm{H}_{2}^{+} + \mathrm{gr}_{}^- \longrightarrow \mathrm{H}_2^{} + 
\mathrm{gr}$ to zero (i.e by preventing the neutralisation of negative 
grains). Preventing the conversion of ${\rm gr}^{2-}$ to ${\rm gr}^-$ by
a similar reaction allows a grain concentration of 
$x_{\mathrm{gr}^{}} \simeq 10^{-13}$ to reproduce the gas--phase chemistry. 
This is because all the neutral grains become and remain negatively charged, 
after which they no longer participate in the chemical evolution.\\
\indent
We now discuss the results of \texttt{model4} with magnesium being 
included with concentration $x_{\mathrm{Mg}^{}} = 10^{-12}$. These 
calculations are presented in figure~\ref{column_model4_c_Mg}. The 
grains are very efficient at sweeping up magnesium, and thermal 
desorption from the grain surfaces is rather inefficient. The result 
is that the active zone with grain concentration $x_{\rm gr}=10^{-12}$ 
and magnesium concentration $x_{\rm Mg}=10^{-12}$ is almost 
indistinguishable from the case with $x_{\rm Mg}=0$ (see 
figure~\ref{column_model1}).\\
\indent
A series of runs were performed to examine the effect of reducing 
% =================================================================  
% figure 14: column density - model4: variation gr
% ================================================================= 
% =================================================================
the grain concentration $x_{\rm gr}$. In particular we are 
interested in defining the grain concentration required to 
reproduce the pure gas--phase chemistry of \texttt{model1} with 
$x_{\rm Mg}=10^{-12}$. The results of these calculations are 
presented in figure~\ref{column_model4_c_Mg}. Reducing $x_{\rm gr}$ 
to $10^{-16}$ results in an active zone whose size is the same as 
that produced by the pure gas--phase chemistry (\texttt{model1}) 
but with $x_{\rm Mg}=0$. In order to obtain an active zone that is 
the same size as that obtained with the gas--phase chemistry 
(\texttt{model1}) with $x_{\rm Mg}=10^{-12}$ we had to reduce the 
grain concentration by more than eight orders of magnitude to 
$x_{\rm gr}=10^{-21}$--$10^{-20}$. The reasons for this are similar 
to those already discussed for \texttt{model4} with $x_{\rm Mg}=0$. 
The equilibrium abundance of free electrons is determined by a 
balance between production and destruction processes. Production 
occurs through photoionisation of H$_2$ only. Destruction is 
dominated by the reaction 
${\rm gr} \; + \; {\rm e^-} \; \rightarrow \; {\rm gr}^-$.
Only a very small number of neutral grains are required 
to reduce the equilibrium electron fraction below that obtained 
by the gas--phase chemistry. When Mg is included in the model, 
the gas--phase chemistry produces a larger electron fraction than 
when $x_{\rm Mg}=0$. Obtaining this enhanced electron fraction means 
that the required depletion of grains is greater when 
$x_{\rm Mg} \ne 0$.\\
\indent
Neutral grains are recycled in this case by the reaction 
${\rm gr}^- \; + \; {\rm Mg}^+ \; \rightarrow \; {\rm gr} \; + \; {\rm Mg}$
as ${\rm Mg}^+$ is the dominant gas--phase ion. By switching off this 
reaction (and the counterpart for gr$^{2-}$ + Mg$^+$) in 
\texttt{model4}, and all reactions involving the adsorption of magnesium 
onto grains, we obtained an active zone whose size was the same as that 
obtained by the pure gas--phase chemistry (\texttt{model1}), but now 
with $x_{\rm gr}=10^{-16}$ rather than $10^{-21}$.\\
\indent
Simply switching off adsorption of magnesium onto grains, but continuing 
to allow Mg$^+$ to neutralise negatively charged grains gr$^-$ resulted 
% ================================================================= 
% figure 15: column density - model4: variation Mg
% ================================================================= 
% =================================================================
in a value of $x_{\rm gr}=10^{-20}$ being required to reproduced the 
gas--phase chemistry. Thus the production of a sizeable active zone in 
the presence of small grains clearly requires either a highly diminished 
population of these small grains, or a means of preventing negatively 
charged grains from being continuously neutralised by positive ions.\\[.5em]
\noindent
\underline{\bf \texttt{model6}/\texttt{model7}}\\[0.5em]
\noindent
We now discuss the results obtained with \texttt{model6} and 
\texttt{model7}. We remind the reader that \texttt{model6} was 
constructed by examining the elements contained in the Sano et 
al. (2000) model (equivalent to our \texttt{model5}), and 
extracting all species from table~\ref{list_species_set}. All 
reactions of the UMIST database are considered that involve these species.
\texttt{model7} was constructed similarly by 
specifying a larger set of elements (H, He, C, N. O, S, Si, Mg, 
Fe).\\
\indent
As a general result we find that the location of the transition 
zone is very similar when comparing \texttt{model4}, \texttt{model6} 
and \texttt{model7} when we use a standard grain concentration of 
$x_{\rm gr}=10^{-12}$. This shows that the grain chemistry is 
dominant in determining the ionisation fraction. Given this fact we 
do not discuss \texttt{model6} any further.\\
\indent
We first consider \texttt{model7} when magnesium is neglected 
from the model. Results for this case are shown in 
figure~\ref{column_model7_c_gr} for a variety of chosen grain 
concentrations. We will discuss cases with $x_{\rm Mg} \ne 0$ 
later on in this section.\\
\indent
For grain concentration $x_{\rm gr}=10^{-12}$ the dead zone is 
significantly larger than obtained with the pure gas--phase 
equivalent \texttt{model3}. Decreasing the grain concentration 
to $x_{\rm gr}=10^{-16}$ resulted in an active zone which is 
the same size as obtained by the pure gas--phase model 
(\texttt{model3}). The reasons for this were discussed at length 
when describing the results from \texttt{model4}, and we do not 
repeat them here. We have run numerical experiments in which 
singly negative--charged grains are prevented from neutralising 
{\em via} recombination reactions with positive ions. In these 
models a grain concentration of $x_{\rm gr} \simeq 10^{-13}$ was 
required to reproduce the size of the active zone obtained in 
the  equivalent pure gas--phase model (\texttt{model3}).\\
\indent
We now discuss results obtained using \texttt{model7} with 
$x_{\rm Mg}=10^{-8}$. These are presented in 
figure~\ref{column_model7_c_Mg}. For a grain concentration 
% =================================================================
% figure 16: 
% =================================================================
% =================================================================
of $x_{\rm gr}=10^{-12}$ we find that the active zone is 
almost identical to that obtained when $x_{\rm Mg}=0$. This 
is because the grains are very effective at sweeping up the 
metal atoms and free electrons. 
As with \texttt{model4}, a reduction of the grain 
concentration to $x_{\rm gr}=10^{-16}$ leads to an active zone 
whose size is very similar to that obtained using a pure 
gas--phase chemistry without metals (\texttt{model3}). A further 
reduction is grain concentration to $x_{\rm gr}=10^{-20}$ is 
required to obtain an active zone of the same size as obtained 
by pure gas--phase chemistry with magnesium abundance 
$x_{\rm Mg}=10^{-8}$. Once again the reasons are the same as 
described for \texttt{model4}, and we do not repeat them here. 
However, it is worth re--emphasising the fact that the active 
zone obtained by the pure gas--phase chemistry of \texttt{model3} 
can be obtained for a smaller grain depletion factor 
($x_{\rm gr}=10^{-16}$) if singly--negative charged grains are 
prevented from forming neutral grains {\em via} recombination 
reactions with positive ions.

%%%%%%%%%%%%%%%%%%%%%%%%%%%%%%%%%%%%%%%%%%%%%%%%%%%%%%%%%%%%%%%%%%%% 
\section{Discussion}
\label{discussion}
\begin{figure}[t]
\includegraphics[width=.50\textwidth]{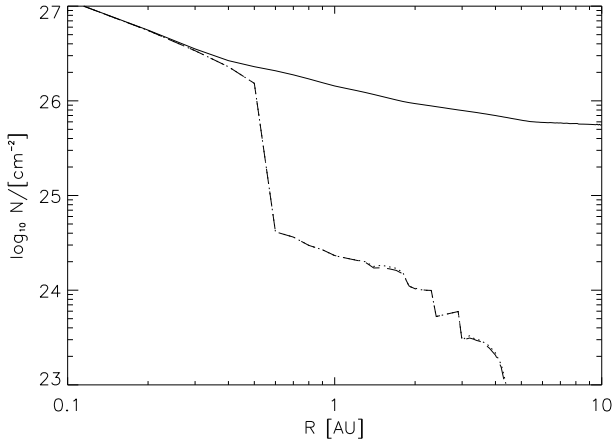}
\caption{- \texttt{model7} - Column densities of the whole disk 
(solid line) and of the active zones (dotted and dashed lines) - 
referring to magnetic Reynolds numbers greater than 100. While 
the dashed line refers to simulation which takes account of all 
types of reactions, the dotted line refers to a simulation where 
all reactions which may form a mantle species were neglected. The 
assumed metal abundances are: 
$x_{\mathrm{Fe}^{}} = 2.74 \times 10^{-9}$ and 
$x_{\mathrm{Mg}^{}} = 1.09 \times 10^{-8}$ while 
$x_{\mathrm{gr}^{}} = 10^{-12}$. The disk parameters are 
$\alpha = 10^{-2}$ and 
$\dot{M} = 10^{-7} \ \mathrm{M_{\odot} yr^{-1}}$.}
\label{figure_discussion_mantle}
\end{figure}

\begin{figure}[t]
\includegraphics[width = .50\textwidth]{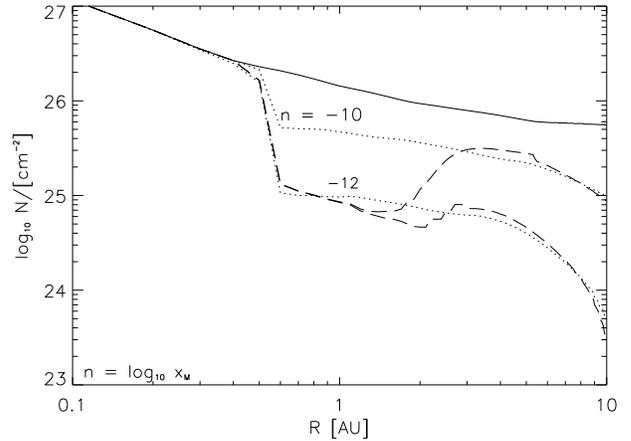}
\caption{Column densities of the whole disk (solid line) and of 
the active zones (dashed and dotted lines) - referring to magnetic 
Reynolds numbers greater than 100 - for two different kinetic models 
of gas--phase chemistry with $x_{\mathrm{Mg}^{}} = 10^{-12}, 10^{-10}$. 
The dashed line refers to the UMIST model (\texttt{model3}) while the 
dotted line refers to the Oppenheimer \& Dalgarno model 
(\texttt{model1}) with modified rate coefficients $\tilde{\alpha}$ and 
$\tilde{\gamma}$. Compared with the values in table~\ref{list_oppenheimer}, 
$\tilde{\alpha}$ and $\tilde{\gamma}$ are increase by 10 and 200, respectively. 
The disk parameters are $\alpha = 10^{-2}$ and 
$\dot{M} = 10^{-7} \ \mathrm{M_{\odot} yr^{-1}}$.}
\label{column_fit}
\end{figure}

\begin{figure}[t]
\includegraphics[width = .50\textwidth]{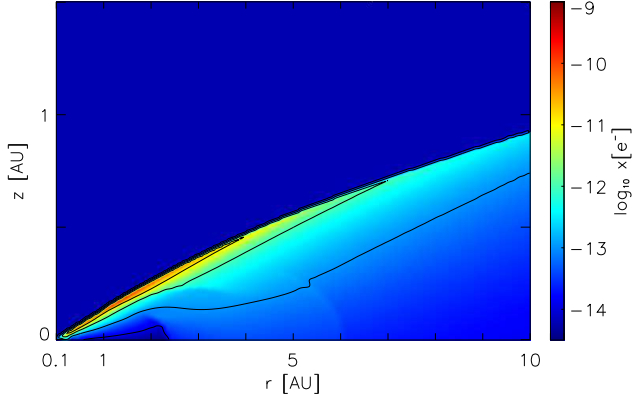}
\caption{This contour plot shows the electron distribution at 
$t=10^5$ yrs for our $\alpha = 10^{-2}$, 
$\dot{M} = 10^{-7} \ \mathrm{M_{\odot} yr^{-1}}$ disk model 
by applying the UMIST model (\texttt{model3}) with 
$x_{\mathrm{Mg}^{}} = 10^{-12}$. The contour lines refer to values 
$x[\mathrm{e}_{}^-]$ of $10^{-14}, 10^{-13}, 10^{-12}$, and $10^{-11}$.}
\label{electron-model3}
\end{figure}

\begin{figure}[t]
\includegraphics[width = .50\textwidth]{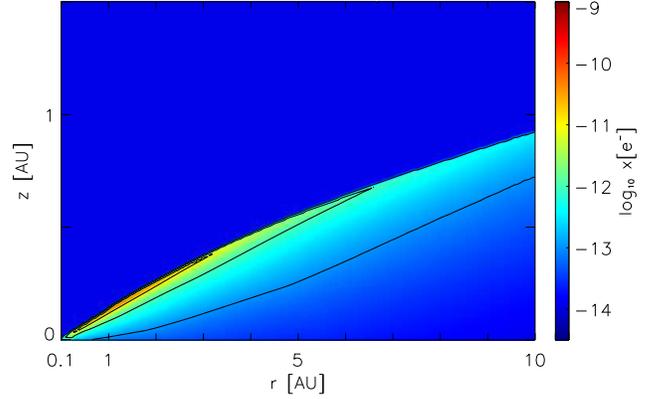}
\caption{This contour plot shows the electron distribution at 
$t=10^5$ yrs for our $\alpha = 10^{-2}$, 
$\dot{M} = 10^{-7} \ \mathrm{M_{\odot} yr^{-1}}$ disk model 
by applying the Oppenheimer \& Dalgarno model (\texttt{model1}) 
with $x_{\mathrm{Mg}^{}} = 10^{-12}$ and modified rate coefficients 
$\tilde{\alpha}$ and $\tilde{\gamma}$. Compared with the values in 
table~\ref{list_oppenheimer}, $\tilde{\alpha}$ and $\tilde{\gamma}$ 
are increase by 10 and 200, respectively. The contour lines refer to 
values $x[\mathrm{e}_{}^-]$ of $10^{-13}, 10^{-12}$, and 
$10^{-11}$.}
\label{electron_fit}
\end{figure}

\begin{figure}[t]
\includegraphics[width=.50\textwidth]{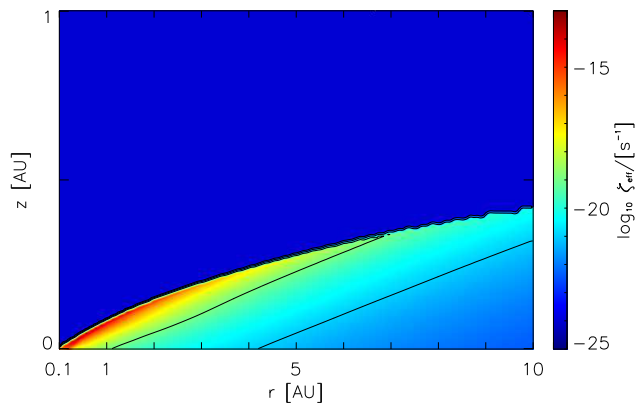}
\caption{The effective X--ray ionisation rate $\zeta_{\mathrm{eff}}$ 
per hydrogen nucleus. The disk parameters are 
$\alpha = 5 \times 10^{-3}$ and 
$\dot{M} = 10^{-8} \ \mathrm{M_{\odot} yr^{-1}}$. The 
contour lines refer to values of $\zeta_{\mathrm{eff}}$: 
$10^{-19}$ and $10^{-21} \ \mathrm{s}^{-1}$.}
\label{figure_zeta2}
\end{figure}

We now discuss issues relating to potential omissions from
% =================================================================
% figure 17: 
% =================================================================
% =================================================================
and extensions to the kinetic models presented in
section~\ref{results}. Apart from the model discussed in 
``Fitting the $x[\rm e_{}^{-}]$ distribution" all other models we 
discuss in this section assumed 
$x_{\mathrm{Mg}^{}} = 1.09 \times 10^{-8}$ and 
$x_{\mathrm{Fe}^{}} = 2.74 \times 10^{-9}$ instead of 
$x_{\mathrm{Mg}^{}} = \mathrm{variable}$ and 
$x_{\mathrm{Fe}^{}} = 0$.\\[.5em]
\noindent
{\sc Radionuclides}\\[0.5em]
\noindent
Sano et al. (2000) and Semenov et al. (2004) included ionisation due 
to the decay of radioisotopes in their models. We have run models with 
this included, using a conservative estimate of the ionisation rate of 
$\zeta_{\mathrm{R}} \approx 6.9 \times 10^{-23} \ \mathrm{s^{-1}}$ 
following Sano et al. (2000). We find that this is not an important 
source of ionisation in our models.\\[.5em]
\noindent
{\sc Nonthermal desorption processes}\\[0.5em]
\noindent
Najita et al. (2001) examined the potential effect of nonthermal 
desorption of grain mantles by X--ray irradiation. We have investigated 
this effect in the simplest possible way, by switching off all reactions 
that involve the adsorption of species onto grain surfaces. It should be 
noted, however, that we still allow electrons to accrete onto grain 
surfaces at the usual rates. We re--ran \texttt{model7} with this 
modification to the reaction network. The results are shown in 
figure~\ref{figure_discussion_mantle}, demonstrating clearly that the 
adsorption of molecular, atomic, and ionic species onto grain surfaces 
plays essentially no role in determining the size of the active zone.\\[.5em]
\noindent
{\sc Fitting the $x[\mathrm{e}_{}^-]$ distribution}\\[0.5em]
\noindent
In section~\ref{results} we computed a variety of
kinetic models for the primary purpose of 
examining the size of the active zone.
Considering first the gas--phase chemistry,
we found significant differences in the
predictions of the simplest model (\texttt{model1}) 
and the most complex model (\texttt{model3}). Future research
% =================================================================
% figure 18: Nonthermal desorption processes
% =================================================================
% =================================================================
in the dynamics of protoplanetary disks will eventually
involve a direct coupling
between magnetohydrodynamic simulations of turbulent disks,
and chemical networks that self--consistently calculate the ionisation
fraction. The task of coupling these two approaches would be
greatly simplified if a chemical network containing
few species could be developed that calculates the ionisation
fraction accurately. For this reason we are motivated
to examine if simple changes to the rate coefficients in the
Oppenheimer \& Dalgarno (1974) network (\texttt{model1})
can lead to agreement with the predicted ionisation
fraction given by the complex UMIST--based model (\texttt{model3}).\\
\indent
The results of our attempts to fit \texttt{model3} by modifying
\texttt{model1} is shown in figure~\ref{column_fit}.
For a metal abundance $x_{\rm Mg}=10^{-12}$ we obtain a good
fit to the size of the active zone by increasing the
rate coefficients $\tilde{\alpha}$ and $\tilde{\gamma}$ by factors of 
10 and 200 compared to the values given in table~\ref{list_oppenheimer}. 
The spatial distribution of the electron fraction obtained using
\texttt{model3} is shown in figure~\ref{electron-model3}
and that obtained using the modified \texttt{model1}
is shown in figure~\ref{electron_fit}. Although there are
small differences, overall these distributions agree well.\\
\indent
The agreement is not so good when we increase the metal abundance
to $x_{\rm Mg}=10^{-10}$. We found that simple modifications
to $\tilde{\alpha}$ and $\tilde{\gamma}$ were unable to produce a 
good overall fit. One is able to fit either the inner ($R \le 2$ AU)
or outer region ($R \ge 2$ AU) quite well, but not both at the same
time (figure~\ref{column_fit} shows a case where we have fitted the 
outer region). This is because the chemistry in \texttt{model1}
is dominated by the metal ion when $x_{\rm Mg}>10^{-12}$, whereas this 
is not the case for \texttt{model3}.\\
\indent
The situation with the gas--grain chemistry is much more simple.
% =================================================================
% figure 19: column density - fitting 
% =================================================================
% =================================================================
For a standard abundance of submicron sized grains, the simple model
(\texttt{model4}) and the complex model (\texttt{model7})
are in extremely good agreement.\\[.5em]
\noindent
{\sc Disk Model}\\[0.5em]
\noindent
In addition to computing the ionisation fraction in disk models with 
$\alpha=10^{-2}$ and ${\dot M}=10^{-7}$ M$_{\odot}$yr$^{-1}$, we also 
computed some cases for which the disk parameters are 
$\alpha=5 \times 10^{-3}$ and ${\dot M}=10^{-8}$ M$_{\odot}$yr$^{-1}$. 
It may be argued that these are closer to the canonical values 
for T Tauri stars. The effective X--ray ionisation rate for this model 
is shown in figure~\ref{figure_zeta2}. The lower mass and density of 
this disk model mean that the ionisation rate is now higher. 
The mass of the disk within the computational domain is now 
about $0.0049 \ \mathrm{M_{\odot}}$ compared with 
$0.0087 \ \mathrm{M_{\odot}}$ for the heavier disk model.\\
\indent 
By applying \texttt{model7} we calculated the ionisation fraction 
for $x_{\mathrm{gr}^{}} = 0$ and  $x_{\mathrm{gr}^{}} = 10^{-12}$. 
The column densities of the active zone are shown in 
figure~\ref{figure_discussion_mdot}. We simply comment that a 
greater percentage of the disk is active when the surface density 
is reduced, but when grains
are absent even a relatively large abundance of metals
is unable to render the disk completely active. 
When grains are present, however, it remains the case 
that only about 1 \% of the matter beyond $R \simeq 1$ AU is 
magnetically active according to our adopted criterion.\\

%%%%%%%%%%%%%%%%%%%%%%%%%%%%%%%%%%%%%%%%%%%%%%%%%%%%%%%%%%%%%%%%%%%%
\section{Summary}
\label{summary}
We have presented calculations of the ionisation fraction
in $\alpha$--disk models using a number of chemical reaction
networks that have appeared in the literature. The primary aims 
are: to compare the predictions of these networks
for the ionisation degree in identical disk models;
to examine the role of gas--grain chemistry in determining
% =================================================================
% figure 20: image electron-model3-typeC 
% =================================================================
% =================================================================
the ionisation fraction in protoplanetary disks;
to examine the level of grain depletion required
to converge toward the pure gas--phase chemistry.
Our main conclusions are:\\
(1). Solving equation~(1) to obtain the electron fraction,
using the approximation that the neutral metal abundance is constant,
leads to an overestimate of the size of the active zone
when heavy metals are included in the gas. \\
(2). In agreement with previous work (Fromang et al. 2002),
we find that the simple Oppenheimer \& Dalgarno (1974)
reaction network predicts globally active disks when the
elemental abundance of metals exceeds $10^{-11}$ and the disk
parameters are ${\dot M}=10^{-7}$ M$_{\odot}$ yr$^{-1}$ and 
$\alpha = 10^{-2}$. \\
(3). A complex gas--phase chemistry drawn from the UMIST data base
predicts larger dead zones than the simple Oppenheimer \& Dalgarno (1974)
model. Whereas disks can be rendered fully active by the addition
of gas--phase heavy metals to the simple model, the complex model
predicts extensive dead zones even for $x_{\rm Mg}=10^{-8}$. \\
(4). A procedure for fitting the results of the complex
gas--phase model, using a modification of the Oppenheimer \& Dalgarno 
scheme, can be successful for specific models, but breaks down
when the abundance of gas--phase metals is increased above 
$x_{\rm Mg}=10^{-12}$. \\
(5). We find that the addition of small grains with concentration
$x_{\rm gr}=10^{-12}$ leads to a dramatic increase in the size of the 
dead zone for all gas--grain chemical models considered. \\
(6). In contrast to the gas--phase chemistry, there is very good agreement
in the predicted size and structure of the active zone
between the different gas--grain chemical networks. This is because the
grains play a dominant role in all gas--grain schemes. \\
(7). A grain depletion factor of $\sim 10^{-4}$ is required to reproduce
the metal--free gas phase chemistry. \\
(8). A grain depletion factor of $\simeq 10^{-8}$ is required
to reproduce the gas phase chemistry including metals.\\
\noindent
Taken at face value, conclusions (7) and (8) suggest that 
% =================================================================
% figure 21: image electron -model1-fit2 
% =================================================================
% =================================================================
efficient growth and settling of small grains will be required 
if gas--phase chemical models are to be used to calculate the 
ionisation structure of protoplanetary disks. Furthermore, this 
requirement must also be met before a significant mass fraction 
of a protoplanetary disk can sustain MHD turbulence. In a recent 
study, Dullemond \& Dominik (2005) calculated the growth and 
settling of grains in laminar and turbulent disks, demonstrating 
that theory predicts the rapid removal of small, submicron sized 
grains. Depletion factors of $< 10^{-6}$ were obtained within 
$\simeq 10^4$ yr, with the consequences for observation and theory 
being: the disk is rendered optically thin to the UV (which has 
potential implications for the chemistry and ionisation fraction); 
the 10 $\mu{\rm m}$ silicate feature is reduced or removed from 
the spectral energy distribution; the mid--IR emission undergoes 
rapid decline. The latter two consequences are at odds with  T 
Tauri disk observations (Dullemond \& Dominik 2005), which maintain 
submicron grain populations for longer than $10^6$ yr. A plausible 
scenario is that an initially quiescent disk undergoes grain 
growth and settling, leading to the development of MHD turbulence 
whose strength is modulated by the gas--phase electron fraction. 
Turbulent motions may contribute to the fragmentation of solids, as 
well as grain growth, so that a population of small grains is maintained 
in the  disk. Feedback between grain size distribution and the strength 
of the turbulence may lead to a quasi--steady state in which the rate 
of growth and fragmentation of solids is controlled by the turbulence, 
whose strength is in turn controlled by the grain size distribution. 
Further work is clearly required to test the validity of such a 
scenario.\\[.5em]
\noindent
In this paper we have concentrated on calculating quasi steady--state 
chemical and ionisation profiles in simple disk models,
ignoring potentially important
effects such as turbulent mixing, diffusion, and a time dependent
X--ray flux. We have already
constructed models including some of these effects, based on the reaction
% =================================================================
% figure 22: zeta2
% =================================================================
% =================================================================
networks that we have described in this paper. These more complex
models are described in Ilgner \& Nelson (2005).

%%%%%%%%%%%%%%%%%%%%%%%%%%%%%%%%%%%%%%%%%%%%%%%%%%%%%%%%%%%%%%%%%%%%
\begin{acknowledgements}
\noindent This research was supported by the
European Community's Research Training Networks Programme under 
contract HPRN-CT-2002-00308, "PLANETS". 
\end{acknowledgements}

%%%%%%%%%%%%%%%%%%%%%%%%%%%%%%%%%%%%%%%%%%%%%%%%%%%%%%%%%%%%%%%%%%%%

%%%%%%%%%%%%%%%%%%%%%%%%%%%%%%%%%%%%%%%%%%%%%%%%%%%%%%%%%%%%%%%%%%%%
\appendix

\section{}
\subsection{List of gaseous components and grain particles}
We considered the following elements: H, He, C, N, O, Mg, Si, S, 
Fe:
% =================================================================
% table: gaseous species
% =================================================================
\begin{table}[htb]
\caption{List of gaseous species including free electrons 
and grain particles with different electric excess charges.}
\begin{center}
\begin{tabular}{lllll}\hline\hline
%1.               & 
   H$_{}^{+}$    &
   H             &
   H$_{2}^{}$    &
   H$_{2}^{+}$   &  
   H$_{3}^{+}$   \\
%2. &   
   He            &
   He$_{}^{+}$   &
   C             &
   C$_{}^{+}$    &
   CH            \\
%3. &
   CH$_{}^{+}$    &
   CH$_{2}^{}$    &
   CH$_{2}^{+}$   &
   N$_{}^{+}$     &
   N              \\
%4. &   
   CH$_{3}^{+}$  &
   NH$_{}^{+}$   &
   CH$_{3}^{}$   &
   NH            &
   CH$_{4}^{+}$  \\ 
%5. &   
   NH$_{2}^{+}$  &
   O             &
   NH$_{2}^{}$   &
   O$_{}^{+}$    &
   CH$_{4}^{}$   \\
%6. &    
   OH            & 
   NH$_{3}^{}$   & 
   NH$_{3}^{+}$  &
   OH$_{}^{+}$   &
   CH$_{5}^{+}$  \\ 
%7. &      
   H$_{2}^{}$O  &
   NH$_{4}^{+}$ & 
   H$_{2}^{}$O$_{}^{+}$ & 
   H$_{3}^{}$O$_{}^{+}$ &  
   Mg  \\ 
%8. & 
   C$_{2}^{}$   &
   C$_{2}^{+}$  & 
   Mg$_{}^{+}$  &    
   C$_{2}^{}$H$_{}^{+}$  &
   C$_{2}^{}$H  \\
%9. &   
   CN    &
   C$_{2}^{}$H$_{2}^{}$ &
   CN$_{}^{+}$ &
   C$_{2}^{}$H$_{2}^{+}$ & 
   HCN   \\
%10. &        
   C$_{2}^{}$H$_{3}^{+}$ &
   HNC   & 
   C$_{2}^{}$H$_{3}^{}$  & 
   HCN$_{}^{+}$  &
   H$_{2}^{}$NC$_{}^{+}$ \\
%11. &   
   Si$_{}^{+}$   &
   HCNH$_{}^{+}$ & 
   CO$_{}^{+}$   & 
   CO    &
   Si    \\ 
%12. &
   N$_{2}^{+}$  &
   N$_{2}^{}$   &
   C$_{2}^{}$H$_{4}^{+}$ & 
   SiH$_{}^{+}$  &
   HCO$_{}^{+}$  \\
%13. &
   HCO   &
   SiH   &
   HN$_{2}^{+}$ &
   NO$_{}^{+}$   &
   H$_{2}^{}$CO$_{}^{+}$ \\
%14. &    
   SiH$_{2}^{}$  &
   NO    &
   SiH$_{2}^{+}$ &
   H$_{2}^{}$CO  &
   H$_{3}^{}$CO$_{}^{+}$   \\ 
%15. & 
   SiH$_{3}^{}$   &
   SiH$_{3}^{+}$  &
   S$_{}^{+}$     &
   CH$_{3}^{}$OH$_{}^{+}$ &
   SiH$_{4}^{}$   \\
%16. &   
   CH$_{3}^{}$OH  &
   O$_{2}^{+}$    & 
   S              &
   SiH$_{4}^{+}$  &
   O$_{2}^{}$     \\
%17. &   
   HS$_{}^{+}$    &
   CH$_{3}^{}$OH$_{2}^{+}$ & 
   HS             &
   SiH$_{5}^{+}$  &
   H$_{2}^{}$S    \\
%18. &   
   H$_{2}^{}$S$_{}^{+}$    &
   H$_{3}^{}$S$_{}^{+}$    &
   C$_{3}^{}$              &
   C$_{3}^{}$$_{}^{+}$     &
   C$_{3}^{}$H    \\
%19. &   
   C$_{3}^{}$H$_{}^{+}$   &
   C$_{3}^{}$H$_{2}^{}$    &
   C$_{3}^{}$H$_{2}^{+}$   &
   C$_{2}^{}$N$_{}^{+}$    &
   CNC$_{}^{+}$    \\
%20. &   
    C$_{3}^{}$H$_{3}^{}$    &
    C$_{3}^{}$H$_{3}^{+}$   &
    C$_{3}^{}$H$_{4}^{}$    &
    SiC$_{}^{+}$    &
    C$_{3}^{}$H$_{4}^{+}$   \\
%21. &
    SiC     &
    C$_{2}^{}$O$_{}^{+}$    &
    HCSi    &
    HC$_{2}^{}$O$_{}^{+}$   &
    HCSi$_{}^{+}$ \\
%22. & 
    CH$_{3}^{}$CN$_{}^{+}$  &
    C$_{3}^{}$H$_{5}^{+}$   &
    CH$_{3}^{}$CN   &
    CH$_{2}^{}$CO    &
    SiN$_{}^{+}$   \\
%23. &
    SiN     &
    CH$_{2}^{}$CO$_{}^{+}$   &
    H$_{4}^{}$C$_{2}^{}$N$_{}^{+}$  &
    CH$_{3}^{}$CO$_{}^{+}$  &
    HNSi    \\
%24. &    
    HNSi$_{}^{+}$   &
    CO$_{2}^{+}$    &
    SiO$_{}^{+}$    &
    CS              &
    SiO     \\
%25. &    
    CO$_{2}^{}$     &
    CS$_{}^{+}$     &
    HCS     &
    HCO$_{2}^{+}$   &
    SiOH$_{}^{+}$   \\
%26. &    
    HCS$_{}^{+}$  &
    NS$_{}^{+}$     &
    NS      &
    H$_{2}^{}$CS    &
    H$_{2}^{}$CS$_{}^{+}$  \\
%27. &    
    H$_{3}^{}$CS$_{}^{+}$   &
    HNS$_{}^{+}$    &
    C$_{4}^{+}$     &
    SO$_{}^{+}$     &
    SO      \\
%28. &    
    C$_{4}^{}$      &
    HSO$_{}^{+}$    &
    C$_{4}^{}$H$_{}^{+}$    &
    C$_{4}^{}$H     &
    C$_{3}^{}$N  \\
%29. & 
    C$_{4}^{}$H$_{2}^{+}$   &
    C$_{3}^{}$N$_{}^{+}$    &
    C$_{4}^{}$H$_{2}^{}$    &
    HC$_{3}^{}$N$_{}^{+}$   &
    HC$_{3}^{}$N    \\
%30. &     
    C$_{3}^{}$O$_{}^{+}$    &
    C$_{3}^{}$O   &
    H$_{2}^{}$C$_{3}^{}$N$_{}^{+}$  &
    HC$_{3}^{}$O$_{}^{+}$   &
    C$_{3}^{}$H$_{2}^{}$O$_{}^{+}$  \\
%31. &    
    H$_{3}^{}$C$_{3}^{}$O$_{}^{+}$  &
    C$_{2}^{}$S$_{}^{+}$    &
    C$_{2}^{}$S     &
    Fe$_{}^{+}$     &
    Fe      \\
%32. &
    HC$_{2}^{}$S$_{}^{+}$   &
    SiS     &
    OCS$_{}^{+}$    &
    SiO$_{2}^{}$    &
    SiS$_{}^{+}$    \\
%33. & 
    OCS  &
    HSiS$_{}^{+}$   &
    HOCS$_{}^{+}$   &
    SO$_{2}^{+}$    &
    SO$_{2}^{}$     \\
%34. &    
    S$_{2}^{}$      &
    HSO$_{2}^{+}$   &
    H$_{4}^{}$C$_{4}^{}$N$_{}^{+}$  &
    H$_{2}^{}$S$_{2}^{+}$   &
    C$_{3}^{}$S     \\
%35. &    
    C$_{3}^{}$S$_{}^{+}$    &
    HC$_{3}^{}$S$_{}^{+}$   &
    C$_{7}^{+}$     &
    e$_{}^{-}$      &
    gr   \\
%36. & 
    gr$_{}^{2-}$    &
    gr$_{}^{-}$     &
    gr$_{}^{+}$     &
    gr$_{}^{2+}$    &
    \\ \hline                                                             
\end{tabular}
\end{center}
\label{list_species_set}
\end{table}
% =================================================================

% =================================================================
% table: evaporation energies:
% =================================================================
\begin{table}[htb]
\caption{Evaporation energies of the considered components 
adsorbed onto grain surfaces. The values are tabulated in 
Hasegawa \& Herbst (1993), if not referred to elsewhere. Values 
marked by $\ast$ and $\dagger$ are listed in Sandford \& 
Allamandola (1993) and Yamamoto et al. (1983), respectively.}
\begin{center}%$_{}^{}$
\begin{tabular}{lrlrlr}\hline\hline
% 1
H                 &  350.0$^{\ }$ &
C$_2^{}$H$_3^{}$  & 1760.0$^{\ }$ &
CO$_2^{}$         & 2860.0$^{\ast}$ \\
% 2 
H$_2^{}$          & 450.0$^{\ }$  &             
H$_2^{}$CO        & 1760.0$^{\ }$ &
C$_4^{}$H$_2^{}$  & 2920.0$^{\ }$ \\
% 3
NH                &  604.0$^{\ }$ &
H$_2^{}$S         & 1800.0$^{\ }$ &
SiH               & 2940.0$^{\ }$ \\
% 4
CH                &  654.0$^{\ }$ &
NS                & 2000.0$^{\ }$ &
HC$_3^{}$N        & 2970.0$^{\ }$ \\
% 5
N$_2^{}$          &  710.0$^{\dagger}$ &
SO                & 2000.0$^{\ }$ &
OCS               & 3000.0$^{\ }$ \\
% 6
O                 &  800.0$^{\ }$ &
S$_2^{}$          & 2000.0$^{\ }$ &
C$_3^{}$S         & 3000.0$^{\ }$ \\ 
% 7
N                 &  800.0$^{\ }$ &
CS                & 2000.0$^{\ }$ &
NH$_3^{}$         & 3075.0$^{\ast}$ \\
% 8
C                 &  800.0$^{\ }$ &          
HCS               & 2000.0$^{\ }$ & 
SiH$_2^{}$        & 3190.0$^{\ }$ \\
% 9
NH$_2^{}$         &  856.0$^{\ }$ &  
C$_3^{}$          & 2010.0$^{\ }$ &  
SiH$_3^{}$        & 3440.0$^{\ }$ \\
% 10
CH$_2^{}$         &  956.0$^{\ }$ &
C$_3^{}$H$_2^{}$  & 2110.0$^{\ }$ &
SO$_2^{}$         & 3460.0$^{\ast}$ \\
% 11
CH$_4^{}$         & 1080.0$^{\dagger}$ &
C$_3^{}$H$_3^{}$  & 2220.0$^{\ }$ & 
SiO               & 3500.0$^{\ }$ \\
% 12
S                 & 1100.0$^{\ }$ &            
H$_2^{}$CS        & 2250.0$^{\ }$ &
SiC               & 3500.0$^{\ }$ \\
% 13
CH$_3^{}$         & 1160.0$^{\ }$ &          
C$_3^{}$H         & 2270.0$^{\ }$ &
HCSi              & 3500.0$^{\ }$ \\
% 14
O$_2^{}$          & 1210.0$^{\ }$ &            
CH$_3^{}$CN       & 2270.0$^{\ }$ &
SiN               & 3500.0$^{\ }$ \\
% 15
C$_2^{}$          & 1210.0$^{\ }$ &          
C$_2^{}$H$_2^{}$  & 2400.0$^{\dagger}$ &
HNSi              & 3500.0$^{\ }$ \\
% 16
CO                & 1210.0$^{\ }$ &          
C$_4^{}$          & 2420.0$^{\ }$ &
SiO$_2^{}$        & 3500.0$^{\ }$ \\
% 17
NO                & 1210.0$^{\ }$ &           
C$_3^{}$H$_4^{}$  & 2470.0$^{\ }$ &
SiH$_4^{}$        & 3690.0$^{\ }$ \\
% 18
OH                & 1260.0$^{\ }$ &         
C$_2^{}$S         & 2500.0$^{\ }$ &
SiS               & 3800.0$^{\ }$ \\
% 19
C$_2^{}$H         & 1460.0$^{\ }$ &                
C$_3^{}$O         & 2520.0$^{\ }$ & 
HCN               & 4170.0$^{\dagger}$ \\
% 20
HS                & 1500.0$^{\ }$ &          
CH$_2^{}$CO       & 2520.0$^{\ }$ &
Fe                & 4200.0$^{\ }$ \\
% 21
CN                & 1510.0$^{\ }$ &           
C$_4^{}$H         & 2670.0$^{\ }$ &
CH$_3^{}$OH       & 4235.0$^{\ast}$ \\
% 22
HCO               & 1510.0$^{\ }$ &            
Si                & 2700.0$^{\ }$ &
H$_2^{}$O         & 4815.0$^{\ast}$ \\
% 23
HNC               & 1510.0$^{\ }$ &             
C$_3^{}$N         & 2720.0$^{\ }$ &
Mg                & 5300.0$^{\ }$  \\ \hline
\end{tabular}
\end{center}
\label{list_evaporation_temperatures}
\end{table}%
% =================================================================

%%%%%%%%%%%%%%%%%%%%%%%%%%%%%%%%%%%%%%%%%%%%%%%%%%%%%%%%%%%%%%%%%%%%
\subsection{Kinetic model of the adsorbed components}
Collisions of gaseous components with grain particles 
may lead to adsorption of the gaseous components onto 
the surfaces of the grain particles. According to the basic 
assumption of kinetics, the adsorbed counterpart gX of the 
gaseous species X is considered as an individual component. 
In addition, the adsorbed species gX is characterized by the 
grain charge itself. Restricting to neutral and simply 
charged grains, the adsorbed species X(gr), X(gr$_{}^{-}$), 
and X(gr$_{}^{+}$) have to be considered as individual 
components so far. However, it becomes important if processes 
between ionized gaseous species, X$_{}^{+}$, and grain particles 
are described by the kinetic model.\\
\indent
Processes onto grain surfaces which involve neutral and 
ionized gaseous species are listed in 
table~\ref{list_mantle_chemistry}. Note, the notation, e.g., 
X(gr$_{}^{-}$) refers to an adsorbed species of a neutral 
gaseous counterpart X onto a grain with simply negative charge. 
In addition be aware that not every reaction listed in 
table~\ref{list_mantle_chemistry} may be applied for each 
gaseous species; cf. for example NH$_{4}^{+}$.\\
\indent
Three different processes are considered: i) adsorption 
processes - reactions 1 to 3 - , ii) desorption processes - 
reactions 6 to 8, and iii) collisional charging processes - 
reactions 4 and 5.\\
% =================================================================
% table: mantle chemistry
% =================================================================
\begin{table}[ht]
\caption{Mantle chemistry}
\begin{center}
\begin{tabular}{r@{\extracolsep{0.1cm}}l@{\extracolsep{0.1cm}}
c@{\extracolsep{0.2cm}}l@{\extracolsep{0.5cm}}
c@{\extracolsep{0.5cm}}l@{\extracolsep{0.2cm}}
c@{\extracolsep{0.2cm}}l@{\extracolsep{0.3cm}}}\hline\hline
1. & 
X            & + & gr$_{}^{-}$    & 
$\longrightarrow$ &
X(gr$_{}^{-}$) &  &  \hspace*{.2cm}\\ 
2. &
X$_{}^{}$      & + & gr           & 
$\longrightarrow$ & 
X(gr)        &   &   \hspace*{.2cm}      \\
3. &
X            & + & gr$_{}^{+}$    & 
$\longrightarrow$ & 
X(gr$_{}^{+}$) &   & \hspace*{.2cm} \\
4. &
X$_{}^{+}$   & +  &   gr$_{}^{-}$ & 
$\longrightarrow$ &  
X & +  & gr \hspace*{.2cm} \\ 
5. &
X$_{}^{+}$     & + & gr           & 
$\longrightarrow$ &  
X(gr$_{}^{+}$) &   & \hspace*{.2cm} \\
6. &
X(gr$_{}^{-}$) &   &              & 
$\longrightarrow$ & 
X            &   &           \hspace*{.2cm} \\
7. & 
X(gr)        &   &              & 
$\longrightarrow$ &  
X            &   &           \hspace*{.2cm} \\
8. & 
X(gr$_{}^{+}$) &   &              & 
$\longrightarrow$ & 
X            &   &           \hspace*{.2cm} \\ \hline
\end{tabular}
\end{center}
\label{list_appendix_mantle_chemistry}
\end{table}%
% =================================================================
\indent
Applying the mass action kinetic deterministic model of 
the reactions listed in table~\ref{list_appendix_mantle_chemistry}, 
one can derive a single ordinary differential equation for 
each component. The number of ordinary differential equations 
can be reduced - which becomes important for systems with a 
large number of gaseous components - if one considers the 
total contributions of the mantle components only. Taking 
% =================================================================
% equation: appendix
% =================================================================
\begin{equation}
$$
\begin{tabular}{%
l@{\extracolsep{0.1cm}}
c@{\extracolsep{0.1cm}}
l@{\extracolsep{0.1cm}}
c@{\extracolsep{0.1cm}}
l@{\extracolsep{0.1cm}}
c@{\extracolsep{0.1cm}}
l@{\extracolsep{0.1cm}}}%
% k_1
\mathsize{\overline{k}_1^{}} & 
\mathsize{\equiv} & 
\mathsize{k_1^{}} &=& 
\mathsize{k_2^{}} &=& 
\mathsize{k_3^{}} \\
% k_2
\mathsize{\overline{k}_2^{}} & 
\mathsize{\equiv} & 
\mathsize{k_4^{}} & &        & &    \\
% k_3
\mathsize{\overline{k}_3^{}} & 
\mathsize{\equiv} & 
\mathsize{k_5^{}} & &        & &    \\
% k_4
\mathsize{\overline{k}_4^{}} & 
\mathsize{\equiv} & 
\mathsize{k_6^{}} &=& 
\mathsize{k_7^{}} &=& 
\mathsize{k_8^{}} 
\end{tabular}
\label{eq2_appendix}
%$$
\end{equation}
into account (cf subsection 3.2) one gets
\begin{eqnarray}
\begin{tabular}{%
l@{\extracolsep{0.1cm}}
c@{\extracolsep{0.1cm}}
l@{\extracolsep{0.1cm}}}%
% d/dt N[X]
\mathsize{
  \frac{d}{dt} n[\mathrm{X}]} & = &
\mathsize{
  \overline{k}_2 n[\mathrm{X}_{}^{+}]n[\mathrm{gr}_{}^{-}] + 
  \overline{k}_4 n[\mathrm{gX}] \ - } \\[.5em]
 & &
\mathsize{       
  \overline{k}_1 n[\mathrm{X}]n_{\mathrm{gr}}^{}} \\[1.0em]
% d/dt N[X+]
\mathsize{
  \frac{d}{dt}n[\mathrm{X}_{}^{+}]}  & = &
\mathsize{ 
  - \ \overline{k}_2 n[\mathrm{X}_{}^{+}]n[\mathrm{gr}_{}^{-}] 
  - \overline{k}_3 n[\mathrm{X}_{}^{+}]n[\mathrm{gr}]} \\[1.0em]
% d/dt N[gX]
\mathsize{
   \frac{d}{dt}n[\mathrm{gX}]} & = &
\mathsize{ 
   \overline{k}_1 n[\mathrm{X}]n_{\mathrm{gr}}^{} +   
   \overline{k}_3 n[\mathrm{X}_{}^{+}]n[\mathrm{gr}] \ -  } \\[.5em]
   & &
\mathsize{
   \overline{k}_4 n[\mathrm{gX}] }  \\[0.5em]
% d/dt N[gr-]
\mathsize{
  \frac{d}{dt}n[\mathrm{gr}_{}^{-}]}  & = &
\mathsize{  
   - \ \overline{k}_2 n[\mathrm{X}_{}^{+}]n[\mathrm{gr}_{}^{-}]} \\[.5em]  
% d/dt N[gr]
\mathsize{
  \frac{d}{dt}n[\mathrm{gr}]}  & = &
\mathsize{  
   \overline{k}_2 n[\mathrm{X}_{}^{+}]n[\mathrm{gr}_{}^{-}] - 
   \overline{k}_3 n[\mathrm{X}_{}^{+}]n[\mathrm{gr}]} \\[.5em]
% d/dt N[gr+]
\mathsize{
  \frac{d}{dt}n[\mathrm{gr}_{}^{+}]}  & = &
\mathsize{
   \overline{k}_3 n[\mathrm{X}_{}^{+}]n[\mathrm{gr}]\ ,}
\end{tabular}
\label{eq3_appendix}
\end{eqnarray}
% =================================================================
with $n[\mathrm{gX}] = n[\mathrm{X}(\mathrm{gr}_{}^{-})] + 
n[\mathrm{X}(\mathrm{gr})] + n[\mathrm{X}(\mathrm{gr}_{}^{+})]$ 
and $n_{\mathrm{gr}}^{} = n[\mathrm{gr}_{}^{-}] + 
n[\mathrm{gr}] + n[\mathrm{gr}_{}^{+}]$. The quantity 
$n[\dots]$ refers to the particle number density per 
cm$_{}^{3}$.

\end{document}